\documentclass[
    aps,
    prapplied,          
    reprint, 
    superscriptaddress,
]{revtex4-2}

\usepackage{placeins}
\usepackage{upgreek}
\usepackage{epstopdf}
\usepackage{amsmath}
\usepackage{physics}
\usepackage{bm}
\usepackage{graphicx}
\usepackage[T1]{fontenc}
\usepackage[colorlinks=true, allcolors=blue]{hyperref}
\usepackage{multirow}

\begin{document}

\title{
Sensitivity threshold defines the optimal spin subset for ensemble quantum sensing
}
\author{Suwan I. Kang}
\thanks{These authors contributed equally}
\affiliation{Department of Physics, Korea Advanced Institute of Science and Technology, Daejeon 34141, South Korea}
\author{Minhyeok Kim}
\thanks{These authors contributed equally}
\affiliation{Department of Physics, Korea Advanced Institute of Science and Technology, Daejeon 34141, South Korea}
\author{Sanghyo Park}
\affiliation{Department of Physics, Korea Advanced Institute of Science and Technology, Daejeon 34141, South Korea}
\author{Heonsik Lee}
\affiliation{Department of Physics, Korea Advanced Institute of Science and Technology, Daejeon 34141, South Korea}
\author{Keunyoung Lee}
\affiliation{Department of Physics, Korea Advanced Institute of Science and Technology, Daejeon 34141, South Korea}
\author{Donggyu Kim}
\email{dngkim@kaist.ac.kr}
\affiliation{Department of Physics, Korea Advanced Institute of Science and Technology, Daejeon 34141, South Korea}
\affiliation{OQT Inc., Daejeon 34051, South Korea}

\begin{abstract}
Inhomogeneous broadening and spatial gradients in control fields inevitably produce large variations in characteristic sensitivity across spin ensembles. We derive an analytic expression for the sensitivity of an inhomogeneous ensemble and introduce a sensitivity threshold that identifies the optimal subset of spins. For both pulsed and continuous-wave magnetometry, the optimal ensembles deliver up to an eightfold improvement over conventional schemes relying on nominally uniform regions of the ensemble. We demonstrate phase-only digital holography to implement the optimal ensembles and show that the measured illumination non-uniformity limits the sensitivity by 14\%. This approach enables optimal quantum sensing in optically scattering environments.
\end{abstract}

\maketitle

\section{Introduction}
Ensembles of atomic spins in solids, such as the electron spins of nitrogen-vacancy (NV) centers in diamonds, serve as versatile quantum sensors~\cite{Chen2017, Glenn2018, Schloss2018, Hayashi2018, Hsieh2019, Marshall2022}. External observables such as magnetic and electric fields globally alter the spin states of the ensemble. Optical~\cite{taylor_high-sensitivity_2008} or electrical~\cite{Hrubesch_Electrical_readout} readout converts these state changes into large metrology signals, enabling precise estimation of the observables. For example, a 1~$\mathrm{mm}^3$ diamond can host billions of NV spins and achieve magnetic field sensitivities of 10~$\mathrm{pT/\sqrt{\mathrm{Hz}}}$ (DC) and 0.1~$\mathrm{pT/\sqrt{\mathrm{Hz}}}$ (AC) under ambient conditions~\cite{Sekiguchi2024, Zheng2020Micro, Fescenko2020}. Moreover, their solid-state form factor allows these spin ensembles to be integrated into diverse environments including biological~\cite{Le_Sage_2013, Davis_2018} and geological samples~\cite{ glenn_geological_imaging_2017}, diamond anvil cells~\cite{ Bhattacharyya2024}, semiconductor devices~\cite{KimCMOS2019}, and other complex media~\cite{DK_QRB_2019}.

The sensitivity in these applications generally scales as $1/\sqrt{N}$ (i.e., the standard quantum limit), where $N$ is the number of spin sensors in the ensemble~\cite{budker_optical_2007}. This scaling, however, assumes that the sensors are identical, a condition rarely met in practice for the following reasons. First, spatial variations in coupling to the local spin bath produce nonuniform coherence times across the ensemble~\cite{bauch_decoherence_2020}. Dynamical decoupling sequences~\cite{viola1999DD, Bar_Gill_2013} suppress this dephasing but are applicable to AC sensing. Second, spatial gradients arise in the control fields used for state preparation and readout. 
For instance, the widely used omega-shaped antenna~\cite{Miller2020Spin, Bolshedvorskii2017,Saha2020} provides 99 \% uniformity over less than 5 \% of its active area~\cite{Opaluch2021OptimizedPlanar}. Robust composite pulses~\cite{Aiello2013Composite} can compensate for these gradients, yet the additional pulses introduce a time overhead that compromises sensitivity~\cite{barry_sensitivity_2020}.  

We derive the sensitivity of ensembles composed of inhomogeneous spin sensors. By accounting for nonuniform control fields and inhomogeneous broadening, we quantify each sensor's contribution to the metrology signal. This analysis introduces a sensitivity threshold that identifies the optimal spin ensemble for ensemble quantum sensing. Contrary to the common assumption that control-field inhomogeneity necessarily limits  the sensitivity, we show that spins within nonuniform regions can improve the ensemble sensitivity beyond that achieved using only nominally uniform regions~\cite{Bauch2018, Arunkumar2021, Yoon2023, Choi2025}. Finally, we demonstrate structured illumination as a practical method for spatially selecting the optimal spin ensemble.

\section{Sensitivity threshold}
\begin{figure}
    \centering
    \includegraphics[width=1\linewidth]{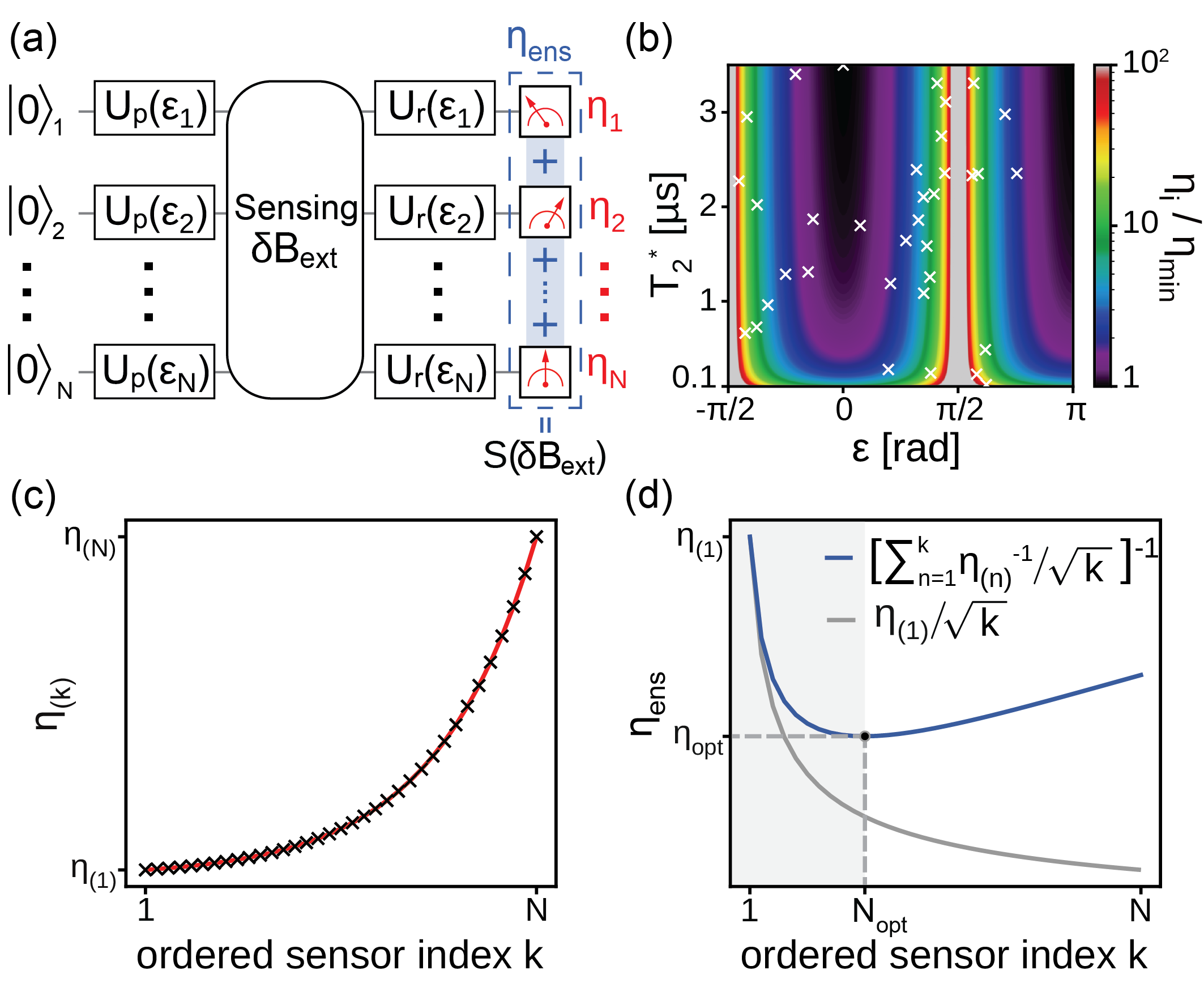}
    \caption{
    \textbf{Sensitivity threshold for quantum sensing with an inhomogeneous spin ensemble.} 
    \textbf{(a)} 
    Global pulses $U_p$ and $U_r$ prepare the ensemble for sensing the phase induced by $\delta B_{\mathrm{ext}}$ and project the accumulated phase onto the spin population, respectively. Spatial variations in the control field and local spin environment produce pulse errors $\epsilon$ and variations in the coherence time $T_2 ^*$, respectively.
    \textbf{(b)}
    Variations in $\epsilon$ and $T_2^*$ produce inhomogeneous characteristic sensitivities $\eta_i$ (cross markers) across the ensemble. Here, $\eta_{\min}$ denotes the minimum characteristic  sensitivity at $\epsilon=0$ and $T_2^* = 3.5$ $\upmu$s.
    \textbf{(c)} The sensors in (b) are ranked according to their characteristic sensitivities. 
    \textbf{(d)} The ensemble sensitivity $\eta_\text{ens}$ reaches its minimum $\eta_{\mathrm{opt}}$ at the optimal ensemble size $N_{\text{opt}}$ (blue line). The dark gray line shows the  $1/\sqrt{N}$ scaling for identical sensors.
    }\label{fig:1-concept}
\end{figure}

Consider an ensemble $\mathcal{S}$ of $N$ atomic spins used to detect a small magnetic field $\delta B_\text{ext}$. The ensemble generates a metrology signal $S(\delta B_\text{ext})$ by repeatedly executing a sensing sequence. For example, a Ramsey-type sensing sequence first applies a global $\pi/2$-pulse ($U_p$) to rotate the optically pumped spins (Fig.~\ref{fig:1-concept}(a)) into the sensing basis~\cite{Degen_Quantum_sensing}. After the spins accumulate phase through the Zeeman interaction with $\delta B_\text{ext}$, a second $\pi/2$-pulse ($U_r$) projects the accumulated phase onto the spin population. Subsequent laser illumination converts the population into  spin-dependent fluorescence $I(\delta B_\text{ext})$ ~\cite{Steiner_2010}. 

Measuring the total ensemble fluorescence $I_\text{ens}=\sum_{i=1}^N I_i(\delta B_\text{ext})$  over a unit time -- during which the sensing sequence is repeated -- produces the metrology signal:
\begin{equation}
    S(\delta B_\text{ext}) \simeq \delta B_\text{ext} \frac{\mathrm{d}I_\text{ens}}{\mathrm{d}B}.
\end{equation}
The ensemble sensitivity $\eta_\text{ens}$, defined as the minimum detectable field at unity signal-to-noise ratio (SNR), is 
\begin{equation}\label{eq:ens_sens_wo_approx}
    \eta_\text{ens} = \Delta I_{\text{noise}}\times\left[\frac{\mathrm{d}I_\text{ens}}{\mathrm{d}B}\right]^{-1}.
\end{equation}
Here, the total measurement noise is $\Delta I_\text{noise}\approx(\sigma_{s}^2 + \sigma_{d}^2)^{1/2}$, where $\sigma_s$ and $\sigma_{d}$ denote the photon shot noise and detector noise, respectively, over the measurement bandwidth. Because the fluorescence contrast between the spin states is small~\cite{Jensen2013}, $\sigma_s\approx \sqrt{I_\text{ens}|_{\delta B_{\text{ext}} = 0}}\approx\sqrt{N I_0}$ where $I_0\simeq I_i(\delta B_\text{ext}=0)$ is the fluorescence from a single spin at $\delta B_\text{ext}=0$.

When the spin sensors are inhomogeneous, they contribute unequally to $S(\delta B_\text{ext})$. Spatial variations in the local noise environment~\cite{bauch_decoherence_2020} produce position-dependent coherence times $T_{c,i}$, while a nonuniform control field introduces pulse errors $\epsilon_i$ in the preparation and readout operations $U_p$ and $U_r$. These inhomogeneities reduce the accumulated phase and spin-dependent fluorescence contrast. In the shot-noise-limited regime, the characteristic sensitivity of the $i$th sensor (Fig. ~\ref{fig:1-concept}(b)) is 
\begin{equation}
    \eta_i(T_{c,i},\epsilon_i)=\sqrt{I_0}\times\left[\frac{\text{d}I_i}{\text{d}B}(T_{c,i}, \epsilon_i)\right]^{-1}.    
\end{equation}

 Nevertheless, because all spins precess at the same Larmor frequency set by $\delta B_{\text{ext}}$, they collectively produce the metrology signal $S(\delta B_{\text{ext}})$. 
Assuming that all $\mathrm{d}I_i/\mathrm{d}B$ have the same sign, the ensemble sensitivity is
\begin{align}
    [\eta_\text{ens}(\mathcal{S})]^{-1} 
        &=  \frac{\sqrt{I_0}} {\sqrt{\sigma_s^2+\sigma_d^2}}\sum_{i=1} ^N \eta_i(T_{c,i},\epsilon_i) ^{-1} \nonumber\\ 
        &\approx \frac{1}{\sqrt{N}}\sum_{i=1} ^N \eta_i(T_{c,i},\epsilon_i) ^{-1}
        \label{eq:ens_sens}
\end{align}
where the second line assumes photon-shot-noise-limited detection, $\sigma_s^2 ~\approx NI_0\gg \sigma_d^2$, as typically achieved for sufficiently large ensembles~\cite{Barry2024, Sekiguchi2024}. For identical sensors, $\eta_i = \eta$, Eq.~(\ref{eq:ens_sens}) reduces to the standard scaling $\eta_\text{ens}=\eta/\sqrt{N}$.

For inhomogeneous sensors, $\eta_\text{ens}$ in Eq.~(\ref{eq:ens_sens}) implies a sensitivity threshold $\eta_{\rm th}$: adding a sensor improves the ensemble sensitivity only when its contribution to the metrology signal $S(\delta B_\text{ext})$ is sufficient to offset the associated increase in photon shot noise. To determine this threshold, we rank the sensors in $\mathcal{S}$ according to $\eta_{(1)}\leq\eta_{(2)}\leq\cdots\leq\eta_{(N)}$, as illustrated in Fig.~\ref{fig:1-concept}(c), and define $\mathcal{S}(k)$ as the subset containing the first $k$ sensors in this ordering. As $k$ increases, $\eta_\text{ens}(\mathcal{S}(k))$ initially decreases, reaches a minimum at $k=N_{\rm opt}$, and then increases as additional sensors contribute insufficient signal relative to their added noise (Fig.~\ref{fig:1-concept}(d)). 
We define the threshold as $\eta_{\rm th}\equiv\eta_{(k=N_{\rm opt})}$, such that the optimal ensemble $\mathcal{S}_{\text{opt}}=\mathcal{S}(N_\text{opt})$ contains the $N_{\text{opt}}$ sensors satisfying $\eta_{(k)}\leq\eta_{\rm th}$.

The ensemble sensitivity in Eq.~(\ref{eq:ens_sens}) does not follow directly from the sum $\sum_i \mathcal{F}_i$, where $\mathcal{F}_i \propto 1/\eta_i^2$ denotes the characteristic Fisher information of the $i$-th sensor~\cite{Degen_Quantum_sensing}. In ensemble sensing, the readout does not resolve individual sensors; instead, the sensors contribute independently to a single collective metrology signal. This collective readout leads to the sensitivity scaling in Eq.~(\ref{eq:ens_sens}), which we also derive explicitly from the Fisher information in the Supplemental Material~\cite{supp}. 

\section{Sensitivity gains}

\begin{figure*}[t]
    \centering
    \includegraphics[width=\linewidth]{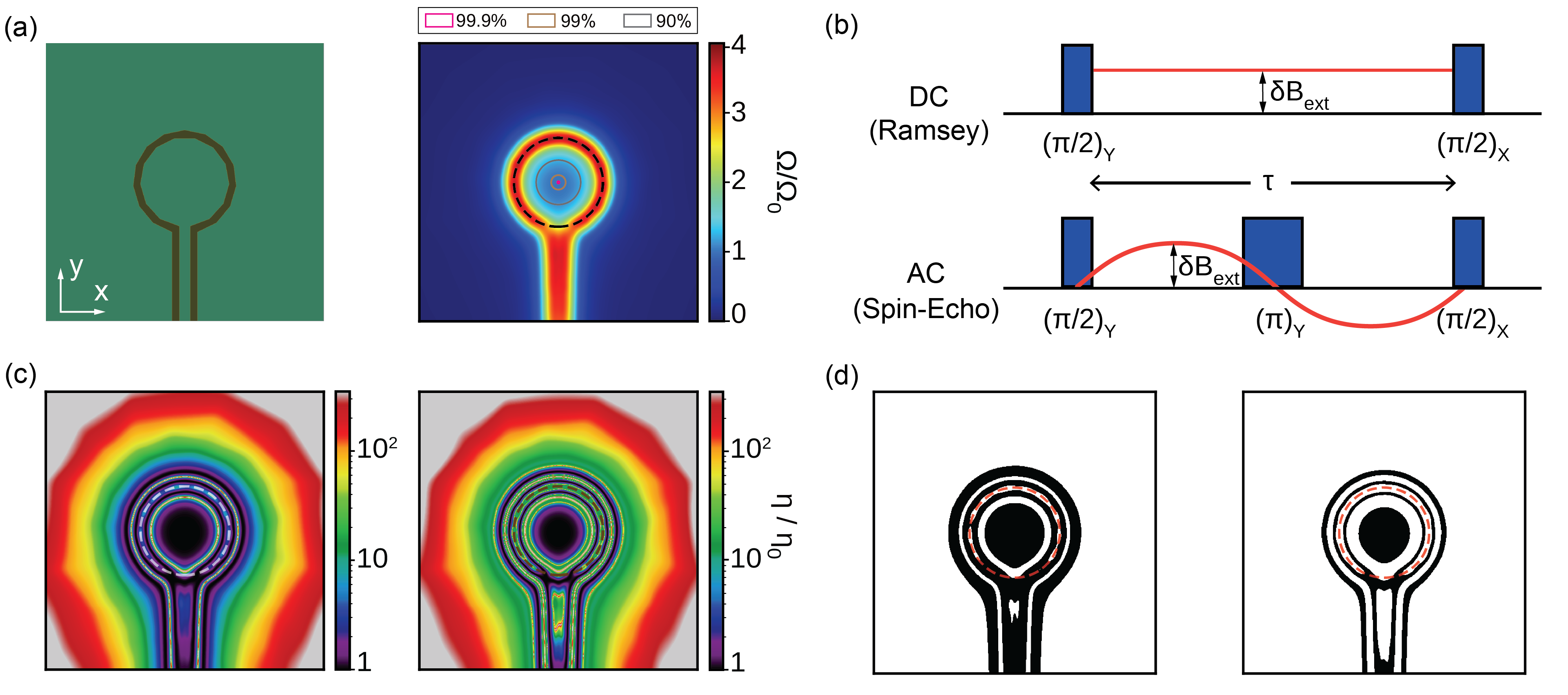}
    \caption{
    \textbf{Optimal spin ensembles for pulsed magnetometry.} 
    \textbf{(a)} Circular loop antenna (left) and simulated normalized Rabi-frequency distribution $\Omega(\mathbf{x})/\Omega_0$ (right),  where $\Omega_0$ is the value at the antenna center. The dashed circle marks the antenna inner boundary, and the solid contours indicate the field uniformities shown in the legend, defined for enclosed area $A$ as $1-\int_A \abs{\Delta\Omega(\mathbf{x})}d\mathbf{x}/\int_A \Omega_0 d\mathbf{x}$. 
    \textbf{(b)} Ramsey and spin-echo sequences for DC and AC  magnetometry;
    \textbf{(c)} Characteristic sensitivities for DC (left) and AC (right) magnetometry, normalized by the sensitivity $\eta_0$ with no pulse error;
    \textbf{(d)} Optimal spin ensembles for DC (left) and AC (right) magnetometry, highlighted in black. 
    }\label{fig:2-pulse}
\end{figure*}

In this section, we identify the optimal spin subsets and quantify their sensitivity gains for an ensemble positioned above the circular loop antenna shown in Fig.~\ref{fig:2-pulse}(a). This geometry is commonly employed to generate a near-uniform control field $\Omega(\mathbf{x})$~\cite{Miller2020Spin, Bolshedvorskii2017, Saha2020}. We assume identical inhomogeneous and homogeneous coherence times, $T_2^*$ and $T_2$, for all spin sensors. We write the local Rabi frequency as $\Omega(\mathbf{x})=\Omega_0+\Delta\Omega(\mathbf{x})$, where $\Omega_0$ is the Rabi frequency at the antenna center.  
For a spin ensemble $\mathcal{S}$, the sensitivity gain $G(\mathcal{S})$ relative to a reference ensemble $\mathcal{R}$ is defined as
 \begin{equation}\label{eq:Gain}
    G(\mathcal{S})= - 20\log_{10}\frac{\eta_{\mathrm{ens}}(\mathcal{S})}{\eta_{\mathrm{ens}}(\mathcal{R})}.     
 \end{equation}
Here, $\mathcal{R}$ contains the spins within the 99\% control-field-uniformity region shown in Fig.~\ref{fig:2-pulse}(a), consistent with the criterion used in previous ensemble-sensing studies~\cite{Rezinkin2024, Chen2022}. Assuming a uniform spin density, the common density factor cancels in the sensitivity ratio, leaving the gain determined by the selected spatial regions of the ensembles. See the Supplemental Material for a detailed derivation~\cite{supp}. 

\subsection{Pulsed magnetometry}
For the Ramsey and spin-echo sequences used in DC and AC magnetometry, respectively (Fig.~\ref{fig:2-pulse}(b)), the characteristic sensitivities of a spin sensor at position $\mathbf{x}$ in the presence of pulse error $\epsilon(\mathbf{x})$ are:
\begin{equation*}%\label{eq:DC-sens}
    \eta^{\text{(DC)}}(\epsilon{(\mathbf{x})}) = 
    \frac{1}{\gamma_e\tau/\sqrt{t_m}}\frac{1}{e^{-(\tau/T_2^*)}}\frac{1}{C\sqrt{n_{\text{avg}}}\cos^2\epsilon{(\mathbf{x})}}
\end{equation*}
and
\begin{align*}
\begin{split}%\label{eq:AC-sens}
    \eta^{\text{(AC)}}&(\epsilon{(\mathbf{x})}) = 
     \frac{\pi}{\gamma_e\tau/\sqrt{t_m}}\frac{1}{C\sqrt{n_{\text{avg}}}}  \\
    & \times \frac{1}{
    \abs{\frac{1}{2}e ^{-(\tau/2T_2^*)}\sin^2(2\epsilon{(\mathbf{x})})-2e^{-(\tau/T_2)}\cos^4\epsilon{(\mathbf{x})}}.
    }
\end{split}
\end{align*}
Here, $\gamma_e$ is the electron gyromagnetic ratio; $C$ is the fluorescence contrast between the spin states; $n_{\text{avg}}$ is the mean number of detected photons per measurement; $\tau$ is the phase-accumulation time; and $t_m$ is the duration of a single measurement. The pulse error $\epsilon(\mathbf{x})=\pi\Delta\Omega(\mathbf{x})/(2\Omega_0)$ is the rotation-angle error of the $\pi/2$-pulse at $\mathbf{x}$. Detailed derivations are provided in the Supplemental Material~\cite{supp}. In the spin-echo sequence, the refocusing-pulse error causes imperfect
refocusing, leaving a residual \(T_2^*\)-dependent contribution to $\eta^{\mathrm{(AC)}}$. Figure~\ref{fig:2-pulse}(c) shows $\eta^\text{(DC)}(\epsilon(\mathbf{x}))$ and $\eta^\text{(AC)}(\epsilon(\mathbf{x}))$ evaluated for the control field $\Omega(\mathbf{x})$ shown in Fig.~\ref{fig:2-pulse}(a).

The characteristic-sensitivity distributions in Fig.~\ref{fig:2-pulse}(c) determine the sensitivity thresholds that define the optimal spin ensembles for Ramsey and spin-echo magnetometry. Ranking the sensitivities and applying Eq.~(\ref{eq:ens_sens}) gives sensitivity thresholds of \(\eta_{\mathrm{th}}^{\text{(DC)}}=2.65\eta_0^{\mathrm{(DC)}}\) and \(\eta_{\mathrm{th}}^{\text{(AC)}}=2.57\eta_0^{\mathrm{(AC)}}\), respectively, where $\eta_0^{\mathrm{(DC)}}$ and $\eta_0^{\mathrm{(AC)}}$ denote the corresponding sensitivities in the absence of pulse error. Further details of the threshold determination are provided in the Supplemental Material~\cite{supp}. 
The more restrictive sensitivity threshold for the spin-echo sequence results from the additional pulse error introduced by the refocusing $\pi$-pulse. When the refocusing-pulse error is removed while retaining the $\pi/2$-pulse error, the AC sensitivity becomes
\begin{align*}
    \eta'^{\text{(AC)}}(\epsilon{(\mathbf{x})}) 
    &= 
    \frac{\pi}{2\gamma_e}\frac{1}{\tau/\sqrt{t_m}}\frac{1}{e^{-(\tau/T_2)}}\frac{1}{C\sqrt{n_{\text{avg}}}\cos^2\epsilon{(\mathbf{x})}} \\
    &\propto \eta^\text{(DC)}(\epsilon(\mathbf{x})),
\end{align*}
showing that the additional pulse-error dependence arises from the nonideal refocusing pulse.

The optimal ensembles $\mathcal{S}_{\mathrm{opt}}^{\text{(DC)}}$ and  $\mathcal{S}_{\mathrm{opt}}^{\text{(AC)}}$ defined by these thresholds are shown in Fig.~\ref{fig:2-pulse}(d). Both extend beyond the central high-uniformity region of the antenna, indicating that spins outside the 99\% control-field-uniformity region can improve the ensemble sensitivity despite their larger pulse errors. Relative to the reference ensemble $\mathcal{R}$, the optimal ensembles provide the sensitivity gains of $G(\mathcal{S}_{\text{opt}}^{\text{(DC)}})=16.83~\mathrm{dB}$ for DC magnetometry and $G(\mathcal{S}_{\text{opt}}^{\text{(AC)}}) = 14.61~\mathrm{dB}$ for AC magnetometry. The smaller gain for the spin-echo sequence is consistent with its more restrictive sensitivity threshold, which reduces the number of sensors included in the optimal ensemble.

\subsection{Continuous-wave (CW) magnetometry}

\begin{figure}
    \centering
    \includegraphics[width=1\linewidth]{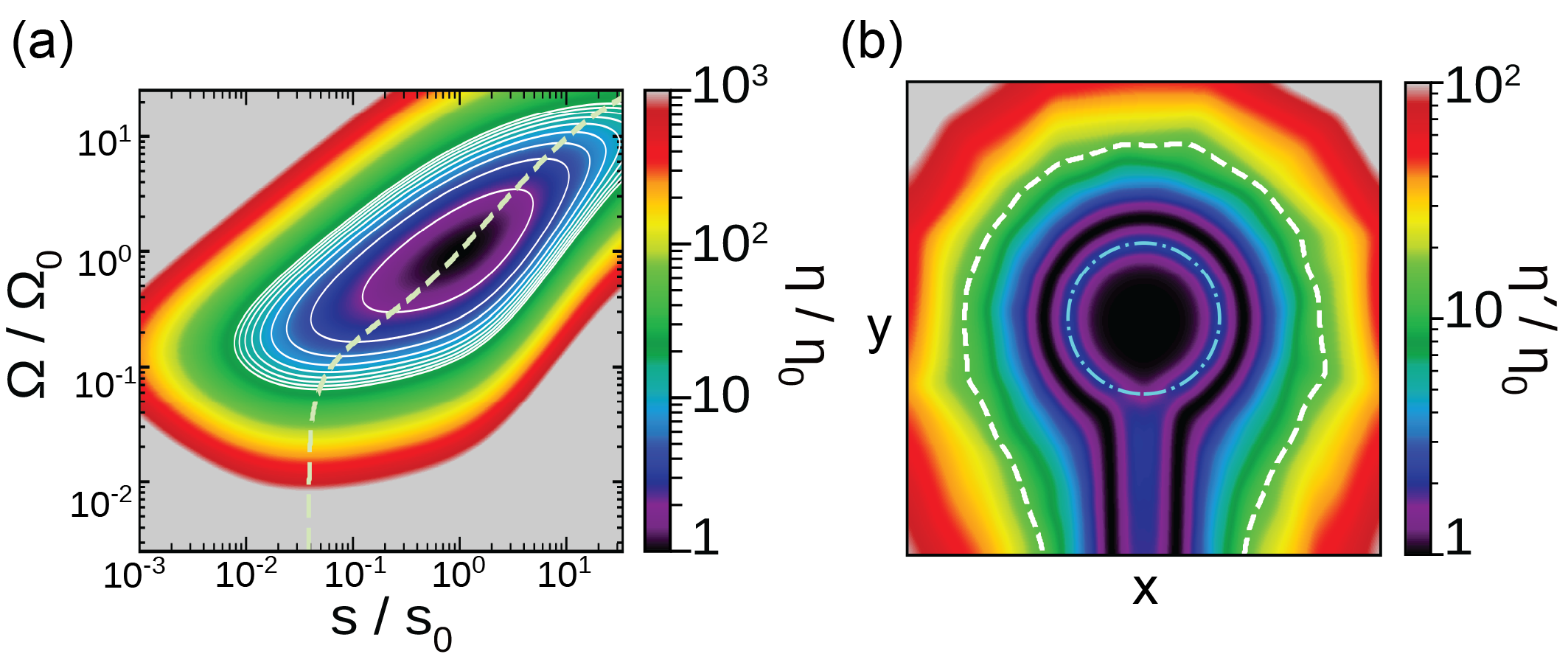}
    \caption{ 
    \textbf{Optimal sensing configuration for CW magnetometry.}
    \textbf{(a)} Normalized characteristic sensitivity $\eta^{\text{(CW)}}(\Omega,s)/\eta^{\text{(CW)}}_0$ of a nitrogen-vacancy center. The dashed line indicates $s'(\Omega)$, which minimizes the characteristic sensitivity at each $\Omega$. Here, $\eta_0$ is the minimum sensitivity of the sensor with $\Omega = \Omega_0$. 
    \textbf{(b)} Spatial distribution of the optimized sensitivity $\eta'^{\text{(CW)}}(\mathbf{x})/\eta^{\text{(CW)}}_0$. The outer white dashed contour indicates the sensitivity threshold defining the optimal ensemble, and the inner cyan dashed circle marks the antenna inner boundary.}
    \label{fig:3. CW-ODMR}
\end{figure}

CW magnetometry continuously generates a metrology signal under near-resonant microwave driving and laser illumination~\cite{Balasubramanian2008}. The external field $\delta B_{\text{ext}}$ is encoded in the steady-state spin population established by the competition between microwave-driven spin transitions and optical pumping \cite{Dreau_avoiding}. This operational simplicity makes CW magnetometry suitable for a wide range of applications~\cite{sengottuvel_wide-field_2022, Xu2023, Zhang2023Optim, Fu2020Sens, Liu2024Fiber}. 

Unlike pulsed magnetometry, a spatially varying control field $\Omega(\mathbf{x})$ in CW magnetometry changes the microwave-driven spin-transition rate, requiring the optical-pumping rate to be adjusted to minimize the characteristic sensitivity~\cite{Dreau_avoiding}. 
Figure~\ref{fig:3. CW-ODMR}(a) shows the characteristic sensitivity $\eta^{\mathrm{(CW)}}(\Omega,s)$ as a function of the Rabi frequency and saturation parameter $s$, which determines the optical-pumping rate. We define $s_0$ as the saturation parameter that minimizes  $\eta_0^{\mathrm{(CW)}}\equiv\eta^{\mathrm{(CW)}}(\Omega_0,s_0)$, the sensitivity of the spin sensor with $\Delta\Omega =0$.
For each $\Omega$, we determine the optimal saturation parameter $s'(\Omega)$ that minimizes $\eta^{\mathrm{(CW)}}(\Omega,s)$. 
Applying $s'(\Omega(\mathbf{x}))$ to the Rabi-frequency distribution yields the optimized characteristic-sensitivity distribution $\eta'^{\mathrm{(CW)}}(\mathbf{x})$, shown in Fig.~\ref{fig:3. CW-ODMR}(b). 
The common microwave frequency is chosen such that $\eta_0^{\mathrm{(CW)}}$ attains its minimum characteristic sensitivity (see the Supplemental Material for details~\cite{supp}).

The sensitivity distribution in Fig.~\ref{fig:3. CW-ODMR}(b) yields a sensitivity threshold of $\eta_{\mathrm{th}}^{\text{(CW)}}=17.23\eta_0^{\mathrm{(CW)}}$, which defines the optimal ensemble $\mathcal{S}_{\mathrm{opt}}^{\mathrm{(CW)}}$ and gives a sensitivity gain of $G(\mathcal{S}_{\mathrm{opt}}^{\mathrm{(CW)}})=18.06~\mathrm{dB}$ (Details are provided in the Supplemental Material~\cite{supp}). The higher normalized threshold relative to Ramsey magnetometry arises because $\Delta\Omega$ in CW magnetometry modifies the microwave-driven spin-transition rate, whose effect is partially compensated by adjusting the optical-pumping condition. In pulsed magnetometry, by contrast, $\Delta\Omega$ directly produces pulse-area errors that reduce the characteristic sensitivity. Adjusting the optical-pumping condition keeps sensors with larger $\Delta\Omega$ below the sensitivity threshold, allowing more spins to be included in the optimal ensemble and yielding a larger sensitivity gain. 

\subsection{Uniformity and Sensitivity Gain}

\begin{figure}[h]
    \centering
    \includegraphics[width=1\linewidth]{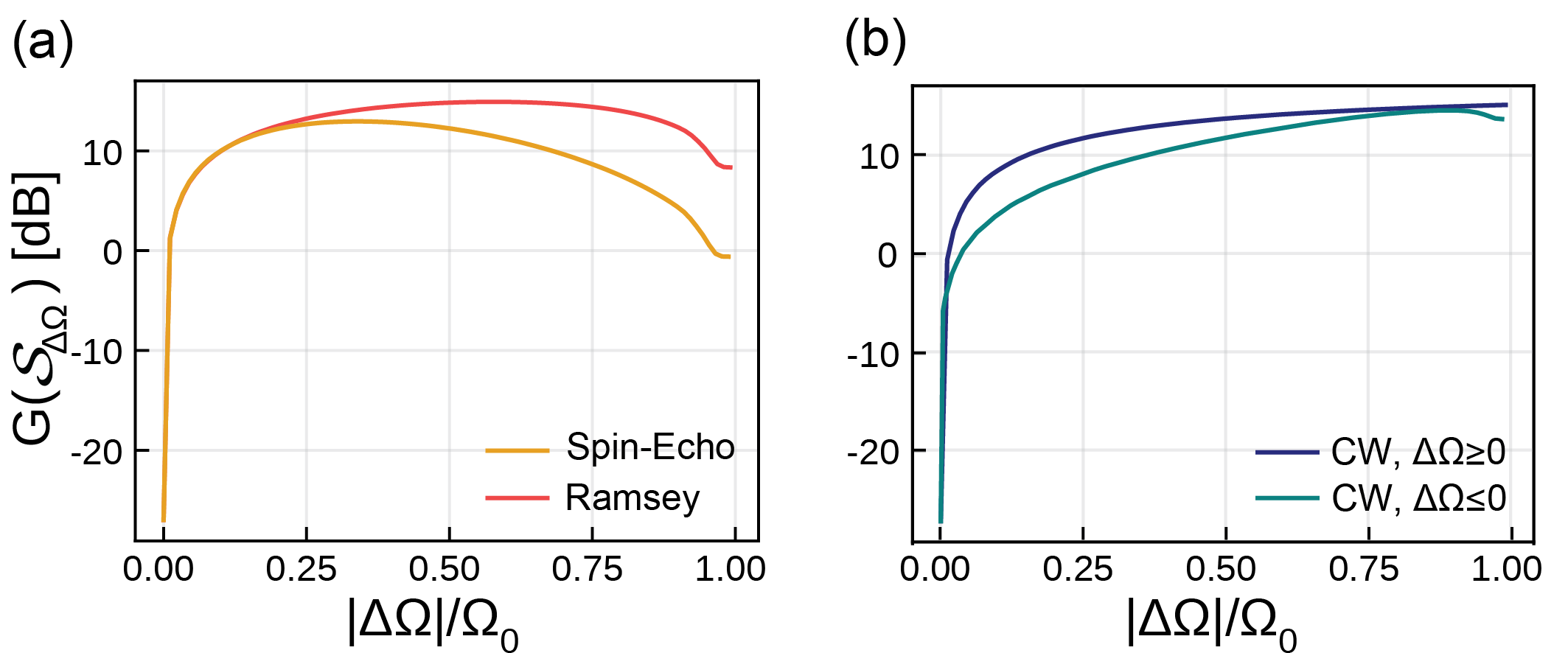}
    \caption{
\textbf{Sensitivity gain versus normalized Rabi-frequency deviations.}
The gain $G(\mathcal{S}_{\Delta\Omega})$ is evaluated relative to the reference ensemble $\mathcal{R}$.
\textbf{(a)} Ramsey and spin-echo magnetometry. For the spin-echo case, sensors with reversed $dI_i/dB$ enter $\mathcal{S}_{\Delta\Omega}$ at $\abs{\Delta\Omega}/\Omega_0 \simeq 0.54$, partially canceling the ensemble metrology signal and reducing $G(\mathcal{S}_{\Delta\Omega})$ below 0 dB as $\abs{\Delta\Omega}/\Omega_0$ approaches unity. See the Supplemental Material for details~\cite{supp};
\textbf{(b)} CW magnetometry for ensembles expanded toward positive or negative $\Delta\Omega/\Omega_0$.
}\label{fig:5-gain}
\end{figure}

To quantify the effect of control-field non-uniformity on ensemble sensitivity, we evaluate the gain $G(\mathcal{S}_{\Delta\Omega})$ (Eq.~(\ref{eq:Gain})), where $\mathcal{S}_{\Delta\Omega}$ represents the spin ensemble as the sensing region is expanded to include locations with progressively larger $\abs{\Delta\Omega}/\Omega_0$. As shown in Fig.~\ref{fig:5-gain}, the gain initially increases as spins outside the high-uniformity region are included and reaches a maximum when additional spins no longer provide sufficient metrology signal to compensate for their shot-noise contribution, placing the optimal ensemble beyond the high-uniformity region without including all available spins. For CW magnetometry with $\Delta\Omega \geq 0$, larger Rabi frequencies improve the characteristic sensitivity of the added spins~\cite{Dreau_avoiding}, allowing the gain to continue increasing over the simulated range (Fig.~\ref{fig:5-gain}(b)).

\section{Implementing optimal spin ensembles}
\begin{figure}
    \centering
    \includegraphics[width=0.5\textwidth]{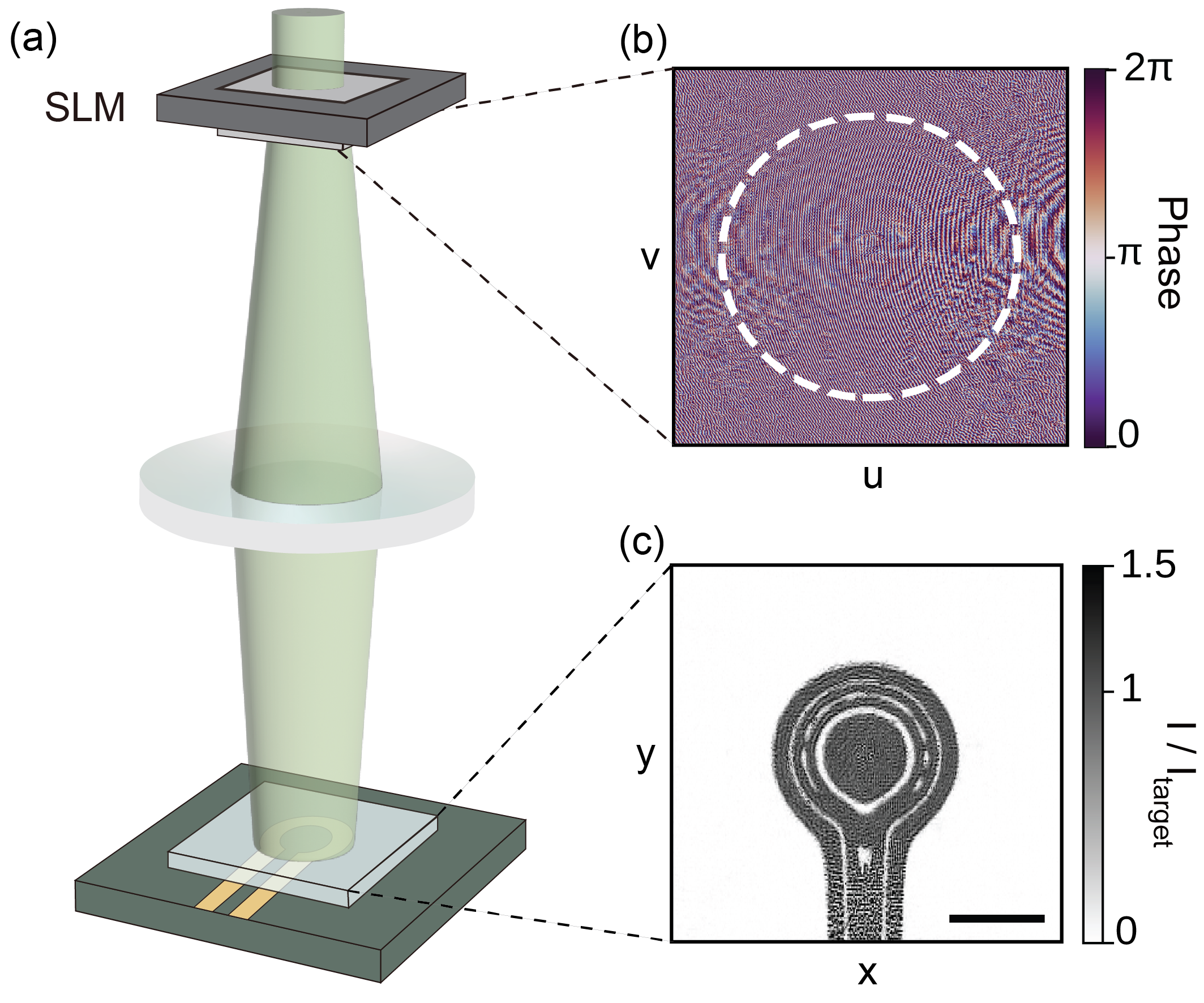}
    \caption{
    \textbf{Structured illumination to implement optimal spin ensembles.}
    \textbf{(a)} Schematic to address the optimal spin ensemble using digital holography with a spatial light modulator (SLM).  
    \textbf{(b)} Phase-only hologram $\phi_\text{DC}(u,v)$; the dashed circle marks the $1/e^2$ diameter of the incident laser beam ($\lambda=532~\text{nm}$).
    \textbf{(c)} Resulting structured illumination pattern $I_\text{DC}(x,y)$. The scale bar in (c) is 1 mm. 
    }
    \label{fig:4-CGH}
\end{figure}

We introduce structured-illumination techniques for implementing optimal ensemble sensing. As illustrated in Fig.~\ref{fig:4-CGH}(a), a spatial light modulator (SLM) shapes the incident laser beam into a target intensity profile $I_{\rm target}(x,y)$ that selectively addresses the optimal spin ensemble. The spin-dependent fluorescence collected under $I_{\rm target}(x,y)$ yields the optimal metrology signal. Figures~\ref{fig:4-CGH}(b) and (c) show the phase-only hologram $\phi_\text{DC}(u,v)$ and the resulting structured illumination $I_\text{DC}(x,y)$ used for DC magnetometry with the circular loop antenna (Fig.~\ref{fig:2-pulse}(a)). The hologram $\phi_\text{DC}(u,v)$ is first generated by an iterative algorithm and subsequently refined through adaptive feedback to correct SLM imperfections. Further methodological details are given in the Supplemental Material~\cite{supp}.

The measured pattern $I_\text{DC}(x,y)$ exhibits a 32.8\% intensity non-uniformity (normalized standard deviation), arising from SLM artifacts such as the fringe effect~\cite{schroff2023accurate, Buske2024}, the cavity effect~\cite{APushkina2020}, and residual aberrations~\cite{Bruce2015}. This non-uniformity introduces a deviation $\delta I_{\text{DC}}(x,y) = I_{\text{DC}}(x,y) - I_{\rm target}(x,y)$ that lowers the SNR of the spin-dependent fluorescence and reduces the sensitivity. A five-level rate-equation model quantifies the resulting sensitivity penalty~\cite{Gupta:16}. 
Specifically, local changes in the readout contrast $C(x,y)$ and the collected photon number $n_{\text{avg}}(x,y)$ in $\eta^\mathrm{(DC)}$ due to $\delta I_{\text{DC}}(x,y)$ are propagated into the ensemble sensitivity in Eq.~(\ref{eq:ens_sens_wo_approx}). 
For the observed 32.8\% non-uniformity, the implemented structured-illumination pattern incurs a 13.4\% sensitivity penalty relative to the target optimal intensity distribution. This small penalty arises because $I_{\text{DC}}(x,y)$ remains close to the saturation required for optimal fluorescence readout~\cite{Gupta:16}. 

\section{Conclusion and Discussion}
In conclusion, we formulate the sensitivity of inhomogeneous spin sensors and derive a sensitivity threshold that identifies the optimal ensemble for ensemble quantum sensing. The optimal ensemble provides a  sensitivity gain by extending the sensing area beyond the nominally uniform region. Our framework can also be extended to other optically detected magnetic resonance (ODMR) readout schemes, such as lifetime-based ODMR~\cite{Strner2021, Lin2025} or time-resolved fluorescence readout \cite{Gupta:16, Chen2019}, once the characteristic sensitivity is defined for the corresponding measurement signal.
We further demonstrate that structured illumination can realize the optimal ensemble with a small sensitivity penalty from illumination non-uniformity. This approach is well suited to ensemble-based quantum sensors requiring high sensitivity, including those used for probing superconductivity~\cite{Bhattacharyya2024}, detecting neuronal signals~\cite{Barry2016}, monitoring electric-vehicle batteries~\cite{Hatano2022}, and inspecting current flows in semiconductor chips~\cite{Garsi2024}. 

Shaping the laser beam to address the optimal spins is formally equivalent to implementing the maximum information state~\cite{Bouchet2021} for spin-based ensemble sensing. This equivalence extends our optimal-ensemble strategy to complex sensor-sample geometries. For example, spin ensembles integrated with multilayer semiconductor circuits \cite{Garsi2024} or embedded in biological media~\cite{Kucsko2013NanometrescaleTI} can require a  tailored illumination that accounts for spatial variations in both spin control and optical detection. This same principle applies to single-spin magnetometry using engineered photonic devices~\cite{Wan2018}, where coherent beam shaping can jointly optimize the spin excitation and photonic boundary conditions. In practice, such an optimal illumination profile could be identified \textit{in} \textit{situ} by iteratively updating the hologram using feedback from spin-based guidestar signals~\cite{DK_QRB_2019}, providing a route to optimal quantum sensing in optically scattering environments. 

\section{Data Availability}
All data necessary to validate the results are fully presented in the article and its Supplemental Material. Additional raw data are available from the corresponding author upon reasonable request.

\clearpage
\bibliographystyle{apsrev4-2}
\bibliography{reference}

\end{document}

\title{
\texorpdfstring{
\textbf{Supplemental Material for \\ ``Sensitivity threshold defines the optimal spin subset for ensemble quantum sensing''}
}{Supplemental Material for "Sensitivity threshold defines the optimal spin subset for ensemble quantum sensing"}
}

\author{Suwan I. Kang}
\thanks{These authors contributed equally}
\affiliation{Department of Physics, Korea Advanced Institute of Science and Technology, Daejeon 34141, South Korea}
\author{Minhyeok Kim}
\thanks{These authors contributed equally}
\affiliation{Department of Physics, Korea Advanced Institute of Science and Technology, Daejeon 34141, South Korea}
\author{Sanghyo Park}
\affiliation{Department of Physics, Korea Advanced Institute of Science and Technology, Daejeon 34141, South Korea}
\author{Heonsik Lee}
\affiliation{Department of Physics, Korea Advanced Institute of Science and Technology, Daejeon 34141, South Korea}
\author{Keunyoung Lee}
\affiliation{Department of Physics, Korea Advanced Institute of Science and Technology, Daejeon 34141, South Korea}
\author{Donggyu Kim} 
\email{dngkim@kaist.ac.kr}
\affiliation{Department of Physics, Korea Advanced Institute of Science and Technology, Daejeon 34141, South Korea}
\affiliation{OQT Inc., Daejeon 34051, South Korea}

\maketitle

\section{Derivation of the Ensemble Sensitivity based on the Fisher Information}
\subsection{Shot-Noise Limited Case}

We first derive the Fisher information $\mathcal{F}$ that a shot-noise limited metrology signal $S(\delta B)$ carries about an external continuous parameter $\delta B$. The signal $S(\delta B)$ may represent a spin-dependent fluorescence of NV centers in the diamond where $\delta B$ is a component of an external magnetic field along the NV center's quantization axis. For the mean $\overline{S(\delta B)}=\lambda(\delta B)$, the likelihood $\mathcal{L}(k|\delta B)$ of observing an outcome  $S(\delta B)=k$ is 
\begin{equation}
    \mathcal{L}(k|\delta B) = \frac{(\lambda(\delta B))^{k}e^{-\lambda(\delta B)}}{k!}
\end{equation}
with its corresponding score function $U(\delta B)$ 
\begin{equation} 
    U(\delta B) \doteq \partial_{\delta B} \log \mathcal{L}(k|\delta B)  = \partial_{\delta B} \lambda(\delta B) \left(\frac{k}{\lambda(\delta B)} - 1\right).
\end{equation}
The Fisher information $\mathcal{F}$ of $S(\delta B)$ is then given by the expectation value of the squared score function $U(\delta B)$:
\begin{equation} 
    \mathcal{F} \doteq  \mathop{\mathbb{E}} [U(\delta B)^2] =  \mathop{\mathbb{E}} \left[(\partial_{\delta B} \lambda(\delta B))^2 \left(\frac{k}{\lambda(\delta B)} - 1\right)^2 \right] = \frac{(\partial_{\delta B} \lambda(\delta B))^2}{\lambda(\delta B)}.
\end{equation}

Now, we consider a metrology signal $S_{\text{ens}}(\delta B)$ generated by a sensor \textit{ensemble}, in which each sensor produces a shot-noise limited signal $\{S_i(\delta B)\}_{i=1, \dots ,N}$. When each sensor is independent and not individually resolvable, the ensemble metrology signal becomes $S_{\text{ens}}(\delta B)=\sum_{i=1}^N S_i(\delta B)$ with mean $\overline{S_{\text{ens}}(\delta B)} = \sum_{i=1}^N\lambda_i(\delta B) \doteq \Lambda(\delta B)$. Since $S_{\text{ens}}(\delta B)$ is also shot-noise limited, the Fisher information $\mathcal{F}_{\text{ens}}$ of $S_{\text{ens}}(\delta B)$ is 
\begin{equation}
    \mathcal{F}_{\mathrm{ens}} = \frac{\left(\partial_{\delta B} \Lambda(\delta B)\right)^2}{\Lambda(\delta B)} = \frac{\left(\sum_{i=1}^N \partial_{\delta B} \lambda_i(\delta B)\right)^2}{\sum_{i=1}^N \lambda_i(\delta B)} \approx \frac{\left(\sum_{i=1}^N \partial_{\delta B} \lambda_i(\delta B)\right)^2}{\sum_{i=1}^N \lambda_i(\delta B=0)}
\end{equation}
for $\delta B \ll 1$. When $ \lambda_i(0) \approx  \lambda_0$ for all $i$, 
\begin{equation}\label{eq:Fisher_ens}
    \mathcal{F}_{\text{ens}} \approx \frac{1}{N\lambda_0}\left(\sum_{i=1}^N\partial_{\delta B}\lambda_i(\delta B)\right)^2.
\end{equation}
Since $ \mathcal{F} = 1/\eta^2$ (where $\eta$ is the sensor sensitivity) for a unit measurement time, 
\begin{equation}
    \partial_{\delta B}\lambda_i(\delta B) = q_i\sqrt{\lambda_i \mathcal{F}_i} \approx q_i\sqrt{\lambda_0}\;\eta_i^{-1}.
\end{equation}
where $q_i =\rm{sgn} \left(\partial_{\delta B}\lambda_i(\delta B)\right)$. 
Substituting this into Eq.~(\ref{eq:Fisher_ens}) gives 
\begin{equation}
    \mathcal{F}_{\text{ens}} = [\eta_\text{ens}(N)]^{-2} = \frac{1}{N\lambda_0}\left(\sum_{i=1}^N\sqrt{\lambda_0}\;q_i\eta_i^{-1}\right)^2,
\end{equation}
which leads to
\begin{equation}
[\eta_\text{ens}(N)]^{-1} = \frac{1}{\sqrt{N}}\sum_{i=1} ^N q_i\eta_i ^{-1}.
\end{equation}
When all $q_i$ take the same value ($q_i = 1$ or $q_i = -1$ for all $i$), this reduces to
\begin{equation}\label{eq:ens_sens}
[\eta_\text{ens}(N)]^{-1} = \frac{1}{\sqrt{N}}\sum_{i=1} ^N \eta_i ^{-1}.
\end{equation}

\subsection{Effect of Detector Noise}\label{section:Detector Noise}
To account for detector noise, the measured fluorescence signal can be written as
\[S_{\text{meas}} = S_{\text{ens}} + \Delta S_{d}\]
where $\Delta S_{d}$ represents a zero-mean detector noise fluctuation. 
Assuming that the photon-shot-noise fluctuations of $S_{\mathrm{ens}}$ and the detector-noise fluctuations $\Delta S_{d}$ are statistically independent, the variance of $S_{\text{meas}}$ becomes $\sigma_{\text{noise}}^2 = \sigma_{s}^2 +\sigma_{d}^2$. Here, $\sigma_{s}^2 = \overline{S_{\text{ens}}(\delta B)} = \Lambda(\delta B)$ denotes the photon shot noise variance, whereas  $\sigma_{d}^2$ denotes the effective detector-noise variance over the measurement bandwidth and is given by
\begin{equation}
    \sigma_{d}^2 = \int_{0}^{\infty}S_{\rm psd}(f)\abs{H(f)}^2 df,
\end{equation}
where $S_{\mathrm{psd}}(f)$ is the detector-noise power spectral density and $H(f)$ is the effective transfer function of the measurement protocol. 
For a sufficiently large mean photon count, the Poisson-distributed ensemble signal can be approximated as a Gaussian random variable. Assuming that the filtered detector-noise fluctuation is also Gaussian, the collective measured signal can be approximated as a Gaussian random variable with mean $\mu(\delta B) = \Lambda(\delta B)$ and variance $V(\delta B) = \Lambda(\delta B) + \sigma_{d}^2$.

The Fisher information of the Gaussian distribution with parameter-dependent mean and variance  is given by 
\[\mathcal{F}(\delta B)= \frac{(\partial_{\delta B} \mu)^2}{V} + \frac{1}{2}\frac{(\partial_{\delta B} V)^2}{V^2}\]
Substituting $\mu= \Lambda(\delta B)$ and $V = \Lambda(\delta B) + \sigma_{d}^2$ gives

\begin{align}
    \mathcal{F}_{\text{ens}}=\frac{(\partial_{\delta B}\Lambda(\delta B))^2}{\Lambda(\delta B) +\sigma_{d}^2} + \frac{1}{2}\frac{(\partial_{\delta B}\Lambda(\delta B))^2}{(\Lambda(\delta B) +\sigma_{d}^2)^2} \approx \frac{(\partial_{\delta B}\Lambda(\delta B))^2}{\Lambda(\delta B) +\sigma_{d}^2} 
\end{align}
where the last approximation is valid in the large-count limit, $\Lambda(\delta B) + \sigma_{d}^2 \gg 1$.
When $ \lambda_i(0) \approx  \lambda_0$ for all $i$,

\begin{align}
    \mathcal{F}_{\text{ens}} \approx \frac{1}{\lambda_0(N + \xi)}\left(\sum_{i=1}^N\partial_{\delta B}\lambda_i(\delta B)\right)^2
\end{align}
where $\xi = \sigma_d^2/\lambda_0$. Following a similar procedure as in the shot noise limited case, the collective sensitivity is modified to
\[[\eta_\text{ens}(N)]^{-1} = \frac{1}{\sqrt{N + \xi}}\sum_{i=1} ^N q_i\eta_i ^{-1}.\]
In the case when the detector noise is negligible compared with the ensemble photon-shot-noise, i.e., $\sigma_d^2 \ll N\lambda_0$ or equivalently $\xi \ll N$ we get the same result assuming all $q_i$ take the same value
\begin{align}
    [\eta_\text{ens}(N)]^{-1}\approx \frac{1}{\sqrt{N}}\sum_{i=1} ^N \eta_i ^{-1}.
\end{align}

\section{Metrology signals for Ramsey Magnetometry}
In this section, we derive the metrology signal for Ramsey magnetometry. The pulse sequences are shown in Fig.~2(b) of the main text.

\subsection{Ramsey Sequence State evolution}
First, we show the state evolution during the pulse sequence in the rotating frame of the MW field oriented perpendicular to the NV symmetry axis $\vec{B}_1(t)=B_1\cos(\omega_0 t -\varphi)\hat{x}$ where $\omega_0 = 2\pi D + \gamma_e B_0$. Here $D$ is the zero-field splitting parameter and $B_0$ is the bias field oriented to the NV symmetry axis. The spin is initialized to $\ket{0}$ state by optical pumping and is followed by a $\pi/2$ pulse $(\varphi = \pi/2)$ $U_p$. Due to the unavoidable amplitude error, the spin does not go to the maximum sensitive state $\ket{+_X} = (\ket{0}+\ket{1})/\sqrt{2}$ perfectly, but it goes to
\begin{align}
\begin{split}
\ket*{\tilde{\psi}(\tau_{\pi/2})} &=\cos(\frac{\frac{\pi}{2}\pm\epsilon}{2})\ket{0}+\sin(\frac{\frac{\pi}{2}\pm\epsilon}{2})\ket{1}\\[5pt]
&=\cos(A)\ket{0}+\sin(A)\ket{1}
\end{split}
\end{align}
where $\displaystyle A = \frac{\pi}{4}\pm\frac{\epsilon}{2}$ and $\epsilon$ is the pulse area error due to the amplitude error. During the interrogation time $\tau$ the spin accumulates the phase induced by $\delta B_{\text{ext}}$ and the state becomes
\begin{align}
\begin{split}
\ket*{\tilde{\psi}(\tau_{\pi/2}+\tau)} &= \exp(-i\frac{1}{2}\gamma_e \delta B_{\text{ext}}\tau\sigma_z)\ket*{\tilde{\psi}(\tau_{\pi/2})}\\[10pt]
&= e^{i\phi/2}\cos(A)\ket{0}+e^{-i\phi/2}\sin(A)\ket{1}
\end{split}
\end{align}
where $\phi=\gamma_e \delta B_{\text{ext}}\tau$.
Finally, a second $\pi/2$ pulse $U_r$ with phase $\varphi$ is applied, and again considering the amplitude error, the state evolves to 
\begin{align}
\begin{split}
\ket*{\tilde{\psi}(\tau_{\pi/2}+\tau+\tau_{\pi/2})} &= e^{i\phi/2}\left[\left[\cos^2(A)-ie^{-i(\varphi+\phi)}\sin^2(A)\right]\ket{0}+\sin(A)\cos(A)\left[e^{-i\phi}-ie^{i\varphi}\right]\ket{1}\right]\\[5pt]
&= \left[\cos^2(A)-ie^{-i(\varphi+\phi)}\sin^2(A)\right]\ket{0}+\sin(A)\cos(A)\left[e^{-i\phi}-ie^{i\varphi}\right]\ket{1}.
\end{split}
\end{align}

\subsection{Ramsey Metrology Signal}
To read out the final state, the state is projected to the basis states $\ket{0}$ and $\ket{1}$ and it is detected via spin dependent fluorescence. To consider this metrology signal we map the basis states $\ket{0}$ and $\ket{1}$ to the coherent states $\ket{\alpha}$, $\ket{\beta}$ respectively which is assumed to be orthogonal and is defined as $\hat{a}\ket{\alpha} = \alpha\ket{\alpha}$ and $\hat{b}\ket{\beta} = \beta\ket{\beta}$ \cite{barry_sensitivity_2020}. Note that for the NV center the zero state fluorescence is greater than the one state fluorescence that is $a=\abs{\alpha}^2 > b=\abs{\beta}^2$. Here, $a$ and $b$ are calculated using the five-level rate-equation model described in Sec.~\ref{section:five-level}. Then the density operator is given as 
\begin{equation}
\hat{\rho} =\abs{\cos^2(A)-ie^{-i(\varphi+\phi)}\sin^2(A)}^2\dyad{\alpha}+\abs{\sin(A)\cos(A)\left[e^{-i\phi}-ie^{i\varphi}\right]}^2\dyad{\beta}
\end{equation}
and therefore the metrology signal is given as follows:
\begin{align}
S(\epsilon, \delta B_{\text{ext}}) &=\ev*{\hat{N}}=\text{Tr}(\hat{\rho}(\hat{a}^\dag\hat{a}+\hat{b}^\dag\hat{b}))\\[10pt]
\label{eq:APP_fluo}&=\frac{(a+b)}{2}+\frac{(a-b)}{2}\sin^2\epsilon -\frac{(a-b)}{2}\cos^2\epsilon\sin(\gamma_e\delta B_{\text{ext}}\tau + \varphi )
\end{align}
Additionally, considering the decreasing contrast due to the dephasing in the interrogation time $\tau$ the metrology signal is given by
\begin{equation}
S(\epsilon, T_2^*, \delta B_{\text{ext}}) =\frac{(a+b)}{2}+\frac{(a-b)}{2}\sin^2\epsilon -\frac{(a-b)}{2}\exp(-\tau/T_2^*)^{p_R}\cos^2\epsilon\sin(\gamma_e\delta B_{\text{ext}}\tau + \varphi ).
\end{equation} 
Here, $T_2^*$ is the inhomogeneous dephasing time, and \(p_R\) is the stretching exponent characterizing the Ramsey coherence envelope; we assume $p_R=1$ throughout. The metrology signal has a maximum slope with respect to $\delta B_{\text{ext}}$ when the phase of the second $\pi/2$ pulse ($U_r$) is $\varphi = 0$. Therefore, the photon-shot-noise-limited sensitivity is given as
\begin{equation}
\eta^{\text{(DC)}} = \frac{1}{\gamma_e\tau/\sqrt{t_m}}\frac{1}{e^{-(\tau/T_2^*)}}\frac{1}{C\sqrt{n_{\text{avg}}}\cos^2\epsilon}
\end{equation}
where $C=(a-b)/(a+b)$ is the contrast, $n_{\text{avg}}=(a+b)/2$ is the mean number of detected photons and $t_m$ is the duration of a single measurement. Here we use small contrast approximation.

\subsection{Ramsey Ensemble Metrology Signal and Ensemble Sensitivity}
In the case of the ensemble consisting of $N$ non-identical sensors, the total metrology signal is given by
\begin{equation}S_{\text{ens}} = \sum_{i=1}^N S(\epsilon_i) = N\frac{(a+b)}{2} + \frac{(a-b)}{2}\sum_{i=1}^N\sin^2\epsilon_i - \frac{(a-b)}{2}e^{-(\tau/T_{2}^{*})}\sin(\gamma_e\delta B_{\text{ext}}\tau+\varphi)\sum_{i=1}^N  \cos^2(\epsilon_i).\end{equation}
From the above equation, we can see that the total fluorescence signal is also sinusoidal for the external DC magnetic field $\delta B_{\text{ext}}$. At the maximal slope point, the photon-shot-noise can be approximated by 
\begin{equation}\delta{S}_{\text{ens}} \approx \sqrt{\frac{(a+b)}{2}N + \frac{(a-b)}{2}\sum_{i=1}^N\sin^2\epsilon_i}\approx \sqrt{N}\sqrt{n_{\text{avg}}}.\end{equation}
where the last approximation is under the assumption of a small contrast $C=(a-b)/(a+b)\ll1$.
Therefore, the field that can be measured is (for simplicity, we shorten $S(\epsilon_i)$ to $S_i$)
\begin{equation}\delta B = \frac{\delta S_{\text{ens}}}{\abs{\frac{\partial\sum_{i=1}^NS_i}{\partial B}}}\approx \frac{\sqrt{N n_{\text{avg}}}}{\sum_{i=1}^{N}\abs{\frac{\partial S_i}{\partial B}}}.\end{equation}
If we approximate the measurement time $t_m$ by the interrogation time $\tau$ and by taking the reciprocal of the above equation, we get 
\begin{equation}\eta^{-1}_{\text{ens}}=\frac{1}{\sqrt{N}}\frac{1}{\sqrt{\tau}}\sum_{i=1}^{N}\frac{\abs{\frac{\partial S_i}{\partial B}}}{\sqrt{n_{\text{avg}}}}\end{equation}
and therefore,
\begin{equation}\label{eq:ramsey_ensemble_sens}
[\eta_\text{ens}(N)]^{-1} = \frac{1}{\sqrt{N}}\sum_{i=1} ^N \eta_i^{-1}.
\end{equation}
which is the same as Eq.~(\ref{eq:ens_sens}).

\section{Metrology signals for spin-echo magnetometry}
In this section, we derive the metrology signal for spin-echo magnetometry. The pulse sequences are shown in Fig.~2(b) of the main text.

\subsection{State Evolution}
Here we show the state evolution in the case of spin-echo sequence for measuring an AC magnetic field  along the NV axis :  $(\delta B_{\text{ext}})\sin(2\pi f_{\text{ac}}t-\varphi_{\text{ac}})$. In the following, $\tau$ denotes the total free-precession time of the spin-echo sequence, with each free-precession interval having a duration $\tau/2$. We consider the resonant spin-echo condition $f_{\text{ac}} = 1/\tau$ and the optimal AC-field phase $\varphi_{\text{ac}} = 0$. Under this condition, the magnitude of the AC-field-induced phase accumulated during each half of the sequence is $\phi_0 = \gamma_e\delta B_{\text{ext}}\tau/\pi$. The state after the first $\pi/2$ pulse is just the same as the Ramsey sequence,
\begin{equation}
\ket*{\tilde{\psi}(\tau_{\pi/2})} = \cos(A_1)\ket{0}+\sin(A_1)\ket{1}
\end{equation}
where $A_1 = \frac{\pi}{4}+\frac{\epsilon}{2}$ and $\epsilon$ is the pulse area error of the $\pi/2$ pulse. During the first interrogation time $\tau/2$, the spin accumulates the magnetic field dependent phase :
\begin{align}
\begin{split}
\ket*{\tilde{\psi}(\tau_{\pi/2}+\tau/2)}&=\exp\left[-\frac{i}{\hbar}\int_0^{\tau/2}\left(\frac{\hbar}{2}\gamma_e \delta B_{\text{ext}}\sin(2\pi f_{\text{ac}}t)\sigma_z+\frac{\hbar}{2}\gamma_e b_j(t)\sigma_z\right)dt\right]\ket*{\tilde{\psi}(\tau_{\pi/2})}\\[10pt]
&=\begin{pmatrix}
\exp(-i\phi_0/2) & 0 \\
0 & \exp(i\phi_0/2)
\end{pmatrix} \begin{pmatrix}
\exp(-i\phi_j/2) & 0 \\
0 & \exp(i\phi_j/2)
\end{pmatrix}\ket*{\tilde{\psi}(\tau_{\pi/2})}\\[10pt]
&=\exp(i(\phi_0+\phi_j)/2)\cos A_1\ket{0} + \exp(-i(\phi_0+\phi_j)/2)\sin A_1\ket{1}.
\end{split}
\end{align}
Here, $b_j(t)$ denotes the zero-mean stochastic fluctuation in the $z$ component of the local magnetic field experienced by the $j$-th NV center, i.e., $\ev*{b_j(t)}=0$.

We conceptually separate this field into a slow or quasi-static component and a dynamic component. The slow component remains approximately constant during a single spin-echo sequence and is refocused by an ideal $\pi$ pulse, whereas the dynamic component is not completely refocused and produces the residual spin-echo decay.
The noise-induced phase accumulated during the first free-precession interval is denoted by 
$\phi_j=\int_0^{\tau/2}\gamma_e b_j(t) dt$. All \(2\times2\) matrices are represented in the ordered basis \((|1\rangle,|0\rangle)\).

After this the $\pi$ pulse is applied to echo out the phase accumulation due to the quasi-static inhomogeneous magnetic field which enlarges the characteristic time of the metrology signal $T_2^*$ to $T_2$. However, due to the unavoidable amplitude error, the pulse area error becomes twice the pulse area error $\epsilon$ of the $\pi/2$ pulse. Considering this amplitude error, the state after the imperfect $\pi$ pulse is given by
\begin{multline}
    \ket*{\tilde{\psi}(\tau_{\pi/2}+\tau/2+\tau_{\pi})} = \left(\cos A_1\cos A_2 - e^{-i(\phi_0+\phi_j)}\sin A_1\sin A_2\right)\ket{0}\\
    +\left(e^{-i(\phi_0+\phi_j)}\sin A_1\cos A_2 +\cos A_1\sin A_2\right)\ket{1}
\end{multline}
where $A_2 = \frac{\pi}{2}+\epsilon$.

For an ideal $\pi$ pulse, the slow or quasi-static phase accumulated during the first free-precession interval is refocused during the second interval. In the presence of a pulse-area error, however, additional signal pathways remain that retain a Ramsey-like coherence factor associated with a single free-precession interval. The properly refocused echo pathway is affected by the residual dynamic-noise coherence characterized by $T_2$.

The second interrogation time comes after the $\pi$ pulse and the phase is accumulated due to the AC field and the inhomogeneous field during $\tau/2$.
\begin{align}
\begin{split}
    &\ket*{\tilde{\psi}(\tau_{\pi/2}+\tau/2 +\tau_{\pi}+\tau/2)}\\
    &=\exp\left[-\frac{i}{\hbar}\int_{\tau/2}^{\tau}\left(\frac{\hbar}{2}\gamma_e (\delta B_{\text{ext}})\sin(2\pi f_{\text{ac}}t)\sigma_z+\frac{\hbar}{2}\gamma_e b_j(t)\sigma_z\right)dt\right]\ket*{\tilde{\psi}(\tau_{\pi/2}+\tau/2+\tau_{\pi})}\\
    &=\left(\cos A_1\cos A_2-e^{-i\phi}\sin A_1\sin A_2\right)\ket{0} + \left(e^{-i(\phi_j+\phi_j')}\sin A_1\cos A_2 +\cos A_1\sin A_2e^{i\phi'}\right)\ket{1}.
\end{split}
\end{align}
Here $\phi = \phi_0+\phi_j$, $\phi'=\phi_0-\phi_j'$ and $\phi_j' = \int_{\tau/2}^{\tau}\gamma_e b_j(t) dt$ which is the accumulated phase due to $b_j(t)$ in the second interrogation time. Finally, a second $\pi/2$ pulse is applied and, again considering the amplitude error, the state evolves to 
\begin{align}
\begin{split}
    &\ket*{\tilde{\psi}(\tau_{\pi/2}+\tau/2+\tau_{\pi}+\tau/2+\tau_{\pi/2})}\\
    &=\left(-ie^{-i(\phi_j+\phi_j')}\sin^2A_1\cos A_2-i\frac{1}{2}\sin (2A_1)\sin A_2 e^{i\phi'} +\cos^2A_1\cos A_2 -e^{-i\phi}\frac{1}{2}\sin (2A_1)\sin A_2\right)\ket{0} \\[10pt]
    & + \left(e^{-i(\phi_j+\phi_j')}\frac{1}{2}\sin (2A_1)\cos A_2 +\cos^2A_1\sin A_2 e^{i\phi'}-i\frac{1}{2}\sin (2A_1)\cos A_2+ie^{-i\phi}\sin^2 A_1\sin A_2\right)\ket{1}.\\
\end{split}
\end{align}

\subsection{Spin-Echo Metrology Signal}
As in the Ramsey case, we map the basis states $\ket{0}$ and $\ket{1}$ to the coherent states $\ket{\alpha}$ and $\ket{\beta}$ respectively and get the density operator of the photons,
\begin{equation}
\hat{\rho} = \abs{\braket*{0}{\tilde{\psi}(\tau_{\pi/2}+\tau/2+\tau_{\pi}+\tau/2+\tau_{\pi/2})}}^2\dyad{\alpha} + \abs{\braket*{1}{\tilde{\psi}(\tau_{\pi/2}+\tau/2+\tau_{\pi}+\tau/2+\tau_{\pi/2})}}^2\dyad{\beta}
\end{equation}
and therefore the metrology signal is given as follows:
\begin{align}
\begin{split}
S &= \expval{\hat{N}}=\text{Tr}(\hat{\rho}(\hat{a}^\dag\hat{a}+\hat{b}^\dag\hat{b}))\\[10pt]
&=\abs{-ie^{-i(\phi_j+\phi_j')}\sin^2A_1\cos A_2-i\sin A_1\cos A_1\sin A_2 e^{i\phi'}+\cos^2A_1\cos A_2-e^{-i\phi}\sin A_1\cos A_1\sin A_2}^2 a\\[10pt]
& \quad +\abs{e^{-i(\phi_j+\phi_j')}\sin A_1\cos A_1\cos A_2 +\cos^2A_1\sin A_2 e^{i\phi'}-i\sin A_1\cos A_1\cos A_2+ie^{-i\phi}\sin^2 A_1\sin A_2}^2 b\\[10pt]
\end{split}
\end{align}

To account for dephasing without assuming a particular microscopic noise distribution, we introduce the effective Ramsey coherence function
\begin{equation}
L_R(t)
\equiv
\left\langle
\exp\left[
i\gamma_e\int_0^t b_j(t')dt'
\right]
\right\rangle
\end{equation}
and the Hahn-echo coherence function
\begin{equation}
L_E(t)
\equiv
\left\langle
\exp\left[
i\gamma_e
\left(
\int_0^{t/2}b_j(t')dt'
-
\int_{t/2}^{t}b_j(t')dt'
\right)
\right]
\right\rangle.
\end{equation}
Here, the brackets denote an average over the relevant stochastic realizations of the local magnetic-field noise. The functions \(L_R(t)\) and \(L_E(t)\) represent effective coherence envelopes of the sensor. These envelopes can be parameterized as stretched exponentials,
\begin{equation}
L_R(t)
=
\exp\left[
-\left(\frac{t}{T_2^*}\right)^{p_R}
\right],
\qquad
L_E(t)
=
\exp\left[
-\left(\frac{t}{T_2}\right)^{p_E}
\right],
\end{equation}
where the stretching exponents depend on the spatial distribution and temporal correlations of the underlying spin bath.
We set \(p_R=p_E=1\) throughout the following analysis. Accordingly,
\begin{equation}
L_R(t)=\exp\left(-\frac{t}{T_2^*}\right),
\qquad
L_E(t)=\exp\left(-\frac{t}{T_2}\right).
\end{equation}
This choice is not intended to specify a particular microscopic decoherence mechanism. More general coherence envelopes, including non-exponential forms can be incorporated by replacing the expressions for \(L_R(t)\) and \(L_E(t)\) without changing the structure of the signal derivation.

Using
\begin{equation}
\left\langle e^{i\phi_j}\right\rangle
=
\left\langle e^{i\phi'_j}\right\rangle
=
L_R(\tau/2)
=
\exp\left(-\frac{\tau}{2T_2^*}\right)
\end{equation}
and
\begin{equation}
\left\langle e^{i(\phi_j-\phi'_j)}\right\rangle
=
L_E(\tau)
=
\exp\left(-\frac{\tau}{T_2}\right),
\end{equation}
the ensemble-averaged spin-dependent fluorescence signal becomes

\begin{multline}
S(\epsilon, T_2^*, T_2, \delta B_{\text{ext}})=\frac{a+b}{2}-\frac{\sin^2\epsilon\cos(2\epsilon)}{2}(a-b)+\frac{a-b}{4}\exp(-\frac{\tau}{2T_2^*})\sin^2(2\epsilon)(\sin(\gamma_e\delta B_{\text{ext}}\tau/\pi)\\
-\cos(\gamma_e\delta B_{\text{ext}}\tau/\pi))-\frac{a-b}{2}\exp(-\frac{\tau}{T_2})\cos^4\epsilon\sin(2\gamma_e\delta B_{\text{ext}}\tau/\pi). \label{eq:echo_signal}
\end{multline}

\noindent
Using this metrology signal the sensitivity can be calculated as follows:

\begin{equation}\frac{\partial S}{\partial B}=\frac{a-b}{4}\exp(-\frac{\tau}{2T_2^*})\frac{\gamma_e\tau}{\pi}\sin^2(2\epsilon)-\frac{a-b}{2}\exp(-\frac{\tau}{T_2})\frac{2\gamma_e\tau}{\pi}\cos^4\epsilon
\label{eq:echo_slope}
\end{equation}

\begin{equation}\delta S=\sqrt{\frac{a+b}{2}-\frac{\sin^2\epsilon\cos(2\epsilon)}{2}(a-b)-\frac{a-b}{4}\exp(-\frac{\tau}{2T_2^*})\sin^2(2\epsilon)}\approx\sqrt{n_{\text{avg}}}\end{equation}

\begin{equation}\eta^{\text{(AC)}} = \delta B\sqrt{t_m} = \frac{\delta S}{\abs{\frac{\partial S}{\partial B}}}\sqrt{t_m} =\frac{\pi}{\gamma_e}\frac{1}{\tau/\sqrt{t_m}}\frac{1}{C\sqrt{n_{\text{avg}}}\abs{\frac{1}{2}\exp(-\frac{\tau}{2T_2^*})\sin^2(2\epsilon)-2\exp(-\frac{\tau}{T_2})\cos^4\epsilon}}\end{equation}
Here $t_m$ is the measurement time. When there is no pulse error, the sensitivity becomes
\begin{equation}\eta_{\epsilon=0}^{\text{(AC)}}=\frac{\pi}{2\gamma_e}\frac{1}{\tau/\sqrt{t_m}}\frac{1}{e^{-\tau/T_2}}\frac{1}{C\sqrt{n_{\text{avg}}}}.\end{equation}

\subsection{Spin-Echo Ensemble Metrology Signal and Ensemble Sensitivity}
In the case of the ensemble consisting of $N$ non-identical sensors, the total fluorescence signal is given as
\begin{align}
\begin{split}
S_{\text{ens}}&=\sum_{i=1}^N S(\epsilon_i)
=\frac{a+b}{2}N - \frac{a-b}{2}\sum_{i=1}^N\sin^2\epsilon_i\cos(2\epsilon_i)+
\frac{a-b}{4}\exp(-\frac{\tau}{2T_2^*})(\sin\phi_0-\cos\phi_0)\sum_{i=1}^N\sin^2(2\epsilon_i)\\[10pt]
&\qquad-\frac{a-b}{2}\exp(-\frac{\tau}{T_2})\sin(2\phi_0)\sum_{i=1}^{N}\cos^4\epsilon_i .
\end{split}
\end{align}
Assuming the contrast $C=(a-b)/(a+b) \ll 1$, the photon-shot-noise can be approximated by
\begin{equation}\delta{S}_{\text{ens}}=\sqrt{\frac{a+b}{2}\left(N-C\sum_{i=1}^N\sin^2\epsilon_i\cos(2\epsilon_i)-\frac{1}{2}C\exp(-\frac{\tau}{2T_2^*})\sum_{i=1}^N\sin^2(2\epsilon_i)\right)}\approx\sqrt{N}\sqrt{n_\text{avg}}.\end{equation}

Because the characteristic sensitivity \(\eta_i\) is defined using the magnitude of the local signal slope, it does not specify whether the \(i\)-th sensor contributes positively or negatively to the collectively
detected signal. We therefore define the local slope-sign factor as
\begin{equation}
q(\mathbf{x})
=
\operatorname{sgn}
\left[
\frac{\mathrm{d}S(\epsilon(\mathbf{x}))}{\mathrm{d}B}
\right],
\end{equation}
and denote \(q_i=q(\mathbf{x}_i)\) for the \(i\)-th sensor at position $\mathbf{x}_i$. Here, \(q(0)\) denotes the slope sign of a pulse-error-free sensor, corresponding to the sensor at the antenna center.

Unlike the Ramsey response, whose local signal slope retains its sign as the pulse-area error varies, the spin-echo signal slope can reverse because the \(T_2^*\)- and \(T_2\)-dependent contributions in
Eq.~\eqref{eq:echo_slope} enter with opposite signs. At \(\epsilon=0\), the slope is determined by the negative \(T_2\)-dependent echo contribution. As the pulse-area error increases, the positive \(T_2^*\)-dependent contribution can become dominant. Consequently, the spin-echo signal slope has the opposite sign to its pulse-error-free value, \(q(\mathbf{x})\neq q(0)\), when
\begin{equation}
\tan^2\epsilon(\mathbf{x})
>
\exp\left(
\frac{\tau}{2T_2^*}
-
\frac{\tau}{T_2}
\right).
\label{eq:echo_sign_reversal}
\end{equation}
The boundaries at which the local signal slope changes sign are therefore
\begin{equation}
\epsilon
=
m\pi
\pm
\tan^{-1}
\left[
\exp\left(
\frac{\tau}{4T_2^*}
-
\frac{\tau}{2T_2}
\right)
\right],
\qquad
m\in\mathbb{Z}.
\label{eq:echo_sign_boundaries}
\end{equation}

The slope signs must be retained when the individual sensor responses are combined. Accordingly, the spin-echo ensemble sensitivity for an arbitrary set of \(N\) sensors is
\begin{equation}
[\eta_{\mathrm{ens}}(N)]^{-1}
=
\frac{1}{\sqrt{N}}
\left|
\sum_{i=1}^{N}
q_i\eta_i^{-1}
\right|.
\label{eq:echo_ens_sens_signed}
\end{equation}
When all local signal slopes have the same sign, \(q_i=q\) for every sensor, Eq.~\eqref{eq:echo_ens_sens_signed} reduces to Eq.~\eqref{eq:ens_sens}. Otherwise, sensors with opposite slope signs contribute oppositely to the collective signal and therefore partially cancel one another. In determining the optimal spin subset in the spin-echo simulations, we therefore restrict the candidate sensors to those satisfying
\begin{equation}
q_i=q(0).
\end{equation} 
This sign restriction is applied only when constructing the optimal spin subset $\mathcal{S}_{\mathrm{opt}}^{(\mathrm{AC})}$(Fig. 2(d) of the main text). In the cumulative expansion $\mathcal{S}_{\Delta\Omega}$ used in Fig. 4(a) of the main text, all spins within the expanded sensing region are included without slope-sign filtering to explicitly show the signal cancellation caused by opposite-slope sensors.

\section{Metrology signals for ODMR-based magnetometry}
\subsection{CW-ODMR}
The optical Bloch equation in the case of CW-ODMR is given as follows~\cite{Dreau_avoiding}:
\begin{align}
\frac{d\sigma_{11}}{dt} &= i\frac{\Omega_R}{2}[\sigma_{10}-\sigma_{01}]-\Gamma_1[\sigma_{11}-\sigma_{00}] - \Gamma_p\sigma_{11}\\[7pt]
\frac{d\sigma_{01}}{dt} &= i[\omega_0-\omega_m]\sigma_{01}-i\frac{\Omega_R}{2}[\sigma_{11}-\sigma_{00}]-\Gamma_2\sigma_{01},
\end{align}
with $\sigma_{11}=1-\sigma_{00}$, $\sigma_{10} = \sigma_{01}^*$. Here, $\Omega_R$ is the Rabi frequency, $\omega_0$ is the resonance frequency, $\omega_m$ is the MW frequency, $\Gamma_1$ is the longitudinal relaxation rate of the populations,  and $\Gamma_p = \Gamma_p^\infty  \frac{s}{1+s}$ denotes the optical spin-polarization rate arising from the intersystem crossing, where \(s\) is the optical saturation parameter and \(\Gamma_p^\infty\) is the maximum polarization rate reached in the high-optical-power limit. $\Gamma_2 = \Gamma_2^* + \Gamma_c$ where $\Gamma_2^*$ is the dephasing rate and $\Gamma_c = \Gamma_c^\infty \frac{s}{1+s}$ describes optical-pumping-induced coherence relaxation, where \(\Gamma_c^\infty\) is its saturation value.

\subsection{ODMR Ensemble Metrology signal and Ensemble Sensitivity}

If there are $N$ non-identical sensors, the total fluorescence is given as
\begin{equation}
S_{\text{ens}} = \sum_{i=1}^N S_i = \sum_{i=1}^N [r_a\sigma_{00}^{\text{st}(i)}+r_b\sigma_{11}^{\text{st}(i)}]\frac{s_i}{1+s_i}
\end{equation}
where $r_a$ and $r_b$ are the fluorescence rates of the $m_s =0$ and $m_s=\pm1$ sublevels, $\sigma_{00}^{\text{st}(i)}$ and $\sigma_{11}^{\text{st}(i)}$ are the steady-state populations of the $i$-th NV sensor and $s_i$ is the saturation parameter applied to the $i$-th NV sensor. In the numerical simulation, the steady-state populations were evaluated at each pixel using the $i$-th Rabi frequency $\Omega_{R,i}$ obtained from the microwave-field map and the $i$-th optimal saturation parameter $s^{'}_i$; 

the resulting $i$-th ODMR spectra were summed to obtain \(S_{\rm ens}\), and the ensemble sensitivity was evaluated at the common microwave frequency chosen to minimize the characteristic sensitivity of the sensor at the antenna center.
Therefore, the ensemble sensitivity is given as
\begin{equation}\eta_{\text{ens}}^{(\text{CW})} = \delta B_{\text{ext}}\sqrt{t_m} = \frac{\sqrt{\sum_{i=1}^NS_i}}{\abs{\frac{\partial\sum_{i=1}^NS_i}{\partial B}}}\sqrt{t_m}\;.\end{equation}

\section{Relation between Sensitivity Gains and Spin Density}
Denoting the ensemble sensitivity obtained using a spin ensemble $\mathcal{S}$ by $\eta_{\text{ens}}(\mathcal{S})$, we defined in the main text the sensitivity gain achieved by $\mathcal{S}$ relative to a reference ensemble $\mathcal{R}$ as $ G(\mathcal{S})=20\log_{10}\left[\eta_{\mathrm{ens}}(\mathcal{R})/\eta_{\mathrm{ens}}(\mathcal{S})\right]$.
Here, we show that this sensitivity gain is independent of the absolute spin density, provided that the spin density is spatially homogeneous and that changing the density does not modify the local spin properties.

Consider a spin ensemble $\mathcal{S}$ occupying a spatial region with area $A_{\mathcal{S}}$. For a spatially uniform areal spin density $\rho$, the number of spins contained in this ensemble is $N_{\mathcal{S}} = \rho A_{\mathcal{S}}$.
In the photon-shot-noise-limited regime, the inverse ensemble sensitivity is given by
\begin{align}
    \eta_{\text{ens}}(\mathcal{S})^{-1} = \frac{1}{\sqrt{N_{\mathcal{S}}}}\sum_{i\in\mathcal{S}}\eta_i^{-1}
\end{align}
where $\eta_i$ is the individual sensitivity of the $i$-th spin. In the continuum limit, the sum over the spins can be expressed in terms of the local sensitivity $\eta(\mathbf r)$ as
\begin{equation}
    \eta_{\text{ens}}(\mathcal{S})^{-1} =\frac{\rho}{\sqrt{N_{\mathcal{S}}}}\int_{\mathcal{S}}\eta^{-1}(\mathbf r) dA = \sqrt{\frac{\rho}{A_{\mathcal{S}}}}\int_{\mathcal{S}}\eta^{-1}(\mathbf r) dA
\end{equation}

Equivalently, defining the spatial average of the inverse local sensitivity over $\mathcal{S}$ as
\begin{equation}
    \expval{\eta^{-1}}_{\mathcal{S}} = \frac{1}{A_{\mathcal{S}}}\int_{\mathcal{S}}\eta^{-1}(\mathbf r) dA
\end{equation}
the ensemble sensitivity can be written as
\begin{equation}
    \eta_{\text{ens}}(\mathcal{S}) = \frac{1}{\sqrt{\rho A_{\mathcal{S}}} \expval{\eta^{-1}}_{\mathcal{S}}}
\end{equation}
Thus, the absolute ensemble sensitivity improves as $\rho^{-1/2}$ with increasing spin density. However, the sensitivity gain compares two ensemble sensitivities evaluated at the same spin density and substituting the above expression into the gain defined in the main text gives
\begin{equation}
    G(\mathcal S) = 20\log_{10}\left[\sqrt{\frac{A_{\mathcal{S}}}{A_{\mathcal{R}}}}\frac{\expval{\eta^{-1}}_{\mathcal{S}} }{\expval{\eta^{-1}}_{\mathcal{R}}}\right]
\end{equation}

The common density factor therefore cancels from the ratio. Consequently, for a spatially homogeneous spin density, the sensitivity gain is independent of the absolute spin density. Instead, it is determined by the selected spatial regions, specifically by their areas and by the distributions of the local spin sensitivities within those regions.

For CW sensing in our work, the inverse-sensitivity relation Eq.~(\ref{eq:ens_sens}) used for pulsed sensing is not directly applicable.

This is because the locally optimized saturation parameter \(s'(\Omega_R)\) changes both the local fluorescence and its magnetic-field slope; consequently, the approximation \(\lambda_i(0)\simeq\lambda_0\) underlying Eq.~(\ref{eq:ens_sens}) does not generally hold for CW sensing.
Nevertheless, its dependence on the spin density and the selected area can be obtained in a similar manner.

We define the local CW-ODMR fluorescence per spin at position $\mathbf r$ as
\begin{equation}
S(\mathbf r, B) = [r_a\sigma_{00}^{\text{st}}(\mathbf r, B)+r_b\sigma_{11}^{\text{st}}(\mathbf r, B)]\frac{s'(\Omega_{R}(\mathbf r))}{1+s'(\Omega_{R}(\mathbf r))}
\end{equation}
where the steady-state populations are evaluated using the local Rabi frequency $\Omega_R(\mathbf r)$, and $s'(\Omega_{R}(\mathbf r))$ is the corresponding locally optimized saturation parameter. 

The total CW-ODMR fluorescence from the spin ensemble $\mathcal S$ is therefore
\begin{equation}
    S_{\text{ens}}(\mathcal S, B) = \rho\int_{A_{\mathcal{S}}}S(\mathbf r, B) dA. 
\end{equation}

Its magnetic-field derivative is correspondingly
\begin{equation}
\frac{\partial  S_{\text{ens}}(\mathcal S, B)}{\partial B} = \rho\int_{A_{\mathcal{S}}}\frac{\partial S(\mathbf r, B)}{\partial B} dA.
\end{equation}
Using the CW ensemble sensitivity defined above, we then obtain
\begin{equation}
    \eta_{\text{ens}}^{(\text{CW})} (\mathcal{S}) = \frac{\sqrt{\rho\int_{A_{\mathcal{S}}}S(\mathbf r, B) dA}}{\rho\abs{\int_{A_{\mathcal{S}}}\frac{\partial S(\mathbf r, B)}{\partial B} dA}}\sqrt{t_m} = \frac{1}{\sqrt{\rho A_{\mathcal{S}}}}\frac{\sqrt{\expval{S}_{\mathcal{S}}}}{\abs{\expval{\partial S/\partial B}_{\mathcal{S}}}}
    \sqrt{t_m}
\end{equation}
where we defined the spatially averaged local ODMR fluorescence and its magnetic-field derivative as
\begin{equation}
    \expval{S}_{\mathcal{S}} = \frac{1}{A_{\mathcal{S}}} \int_{A_{\mathcal{S}}} S(\mathbf r, B) dA
\end{equation}
and
\begin{equation}
\expval{\partial S/\partial B}_{\mathcal{S}} = \frac{1}{A_{\mathcal{S}}}\int_{A_{\mathcal{S}}}\frac{\partial S(\mathbf r, B)}{\partial B} dA.
\end{equation}

This expression shows that the absolute CW ensemble sensitivity depends on the total number of spins, $N_{\mathcal{S}} = \rho A_{\mathcal{S}}$, and improves as $\rho^{-1/2}$ when increasing the spin density does not modify the local ODMR response. Using the sensitivity gain defined in the main text, the ratio between the ensemble sensitivity using spin ensemble $\mathcal S$ and a reference ensemble $\mathcal{R}$ becomes
\begin{equation}
    \frac{\eta_{\text{ens}}^{(\text{CW})} (\mathcal{R})}{\eta_{\text{ens}}^{(\text{CW})} (\mathcal{S})} = \sqrt{\frac{A_{\mathcal{S}}}{A_{\mathcal{R}}}} \frac{\sqrt{\expval{S}_{\mathcal{R}}}\Big/\abs{\expval{\partial S/\partial B}_{\mathcal{R}}}}{\sqrt{\expval{S}_{\mathcal{S}}}\Big/\abs{\expval{\partial S/\partial B}_{\mathcal{S}}}}
\end{equation}

The common spin-density factor therefore cancels from the sensitivity ratio. Consequently, the CW sensitivity gain is also independent of the absolute spin density and is determined by the selected areas and the spatial distributions of the local CW-ODMR fluorescence and its magnetic-field response within those areas.

\section{Time-resolved spin-dependent fluorescence}
\subsection{Five-level Rate-Equation model for \texorpdfstring{$\text{NV}^{-}$}{NV-}} \label{section:five-level}

\begin{figure}[t]
    \centering
    \includegraphics[width=0.5\textwidth]{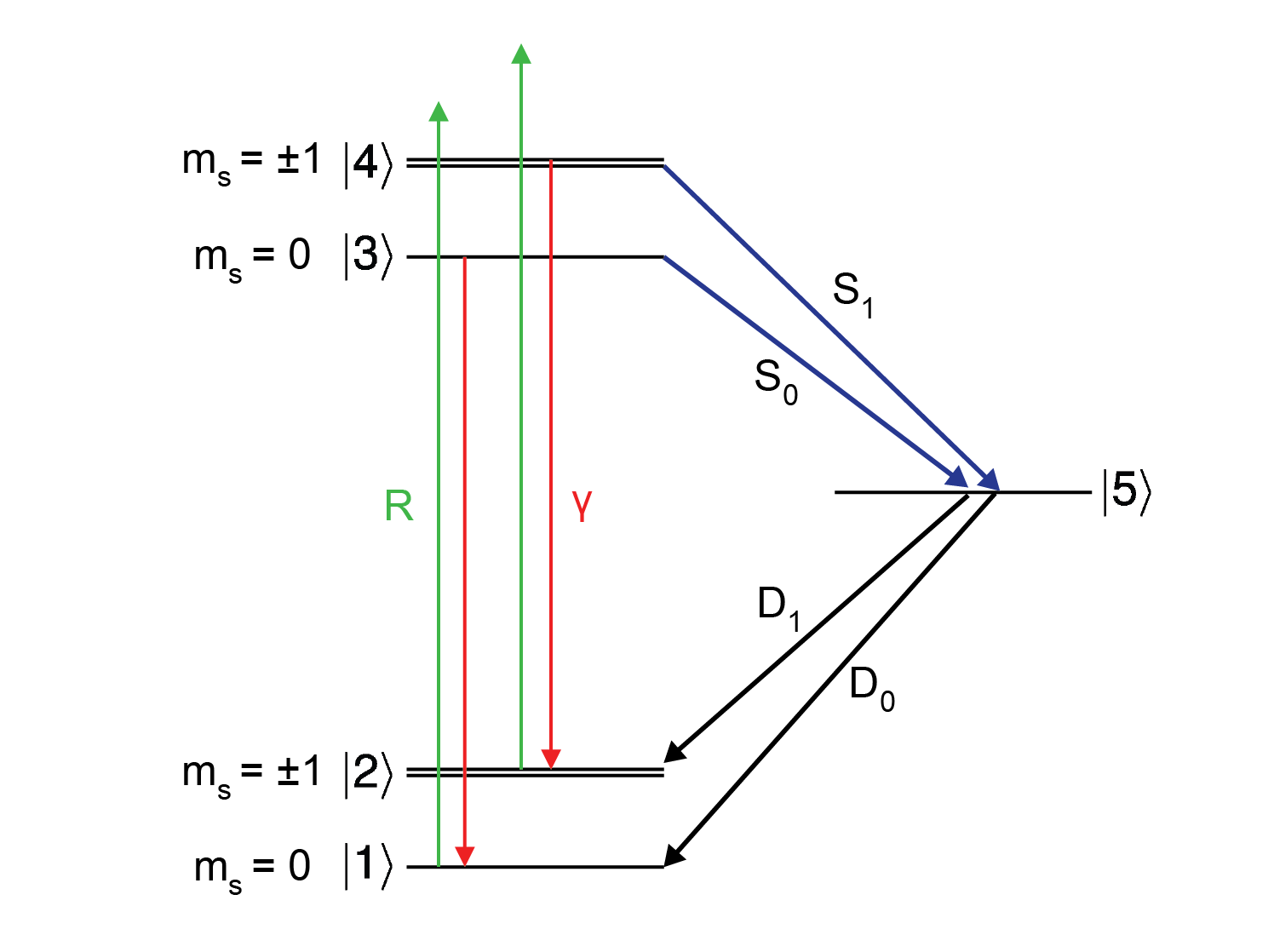}
    \caption{
    Simplified five-level model for $\mathrm{NV}^{-}$
    }
    \label{fig:five_level}
\end{figure}

The time-resolved spin-dependent fluorescence can be obtained by solving the optical rate-equation. If we treat the two singlet states of the $\text{NV}^{-}$ center as one state we can use a five-level optical rate-equation model to describe the $\text{NV}^{-}$ spin-photon dynamics (Fig.~\ref{fig:five_level}). The five-level rate-equation for NV$^-$ is given as follows:
\begin{align}
\frac{dN_1}{dt} &= \gamma N_3 + D_0N_5 - R N_1\\[7pt]
\frac{dN_2}{dt} &= \gamma N_4 + D_1N_5 - R N_2 \\[7pt]
\frac{dN_3}{dt} &=-(\gamma+S_0)N_3 + R N_1\\[7pt]
\frac{dN_4}{dt} &=-(\gamma+S_1)N_4 + R N_2 \\[7pt]
\frac{dN_5}{dt} &=S_0N_3 + S_1N_4 - D_0N_5 - D_1N_5
\end{align}
Here, $N_i$ is the population of the $i$-th state that satisfies $\sum_i N_i = 1$, $R$ is the optical pumping rate that is related to the intensity of the 532 nm green laser, $\gamma$ is the radiative decay rate. $D_0$, $D_1$, $S_0$, $S_1$ are transition rates depicted in Fig.~\ref{fig:five_level}. We used the transition rate values of NV1 reported in Ref.~\cite{Gupta:16}.

\begin{table}[b]
\centering
\caption{Transition rates used in the five-level rate-equation model~\cite{Gupta:16}.}
\label{tab:transition_rate}
\begin{tabular}{llc}
\hline\hline
Parameter & Symbol & Value (MHz) \\
\hline
Optical pumping rate
& $R$
& $1.5R_{\mathrm{sat}} = 41.07$ (Sec.~\ref{sec: saturation intensity})
\\

Radiative decay rate
& $\gamma
$
& $67.4$
\\

Intersystem-crossing rate from $\ket{3}$
& $S_0$
& $9.9$
\\

Intersystem-crossing rate from $\ket{4}$
& $S_1$
& $91.6$
\\

Metastable decay rate to $\ket{1}$
& $D_0$
& $4.83$
\\

Metastable decay rate to $\ket{2}$
& $D_1$
& $2.11$
\\
\hline\hline
\end{tabular}
\end{table}

By solving the five-level rate-equation given above, we can find how the population evolves as a function of time and the time-resolved spin-dependent fluorescence rate can be calculated by $\gamma(N_3 + N_4)$. Therefore, the mean number of photons emitted from the NV can be obtained by
\begin{equation}\label{eq:mean_num_photon}
n = \int_0^{t_{\text{readout}}} \gamma(N_3(t) + N_4(t)) dt    
\end{equation}
where $t_{\text{readout}}$ is the readout time gate width and $n$ is the mean number of photons collected during $t_{\text{readout}}$. 

The photon counts \(a\) and \(b\) were calculated by solving the five-level rate-equations with the initial conditions
\begin{equation}
(N_1,N_2,N_3,N_4,N_5)=(1,0,0,0,0)    
\end{equation}
and
\begin{equation}
(N_1,N_2,N_3,N_4,N_5)=(0,1,0,0,0),
\end{equation}
respectively, corresponding to the spin being initialized in the \(m_s=0\) and \(m_s=\pm1\) ground states at the onset of optical readout. All excited-state and singlet-state populations were initially set to zero. The resulting time-resolved fluorescence was integrated over the readout window $t_{\mathrm{readout}}$, which was optimized by maximizing $C^2 n_{\mathrm{avg}}$, where $C$ and $n_{\mathrm{avg}}$ denote the spin-dependent fluorescence contrast and the mean detected photon count, respectively.

\subsection{Optical pumping rate for saturation intensity} \label{sec: saturation intensity}
The population in the steady state can be found by setting the derivatives of the rate-equation to zero and using the condition $\sum_i N_i = 1$. Then the steady state solution is given as follows:
\begin{align}
N_1^{\text{st}} &= \frac{1}{R}\left(\frac{\gamma D_0}{S_0}+D_0\right)N_5^{\text{st}}\\[7pt]
N_2^{\text{st}} &= \frac{1}{R}\left(\frac{\gamma D_1}{S_1}+D_1\right)N_5^{\text{st}}\\[7pt]
N_3^{\text{st}} &= \frac{D_0}{S_0}N_5^{\text{st}}\\[7pt]
N_4^{\text{st}} &= \frac{D_1}{S_1}N_5^{\text{st}}\\[7pt]
N_5^{\text{st}} &= \frac{1}{\frac{D_0+D_1}{R}+\frac{(\gamma+R)D_0}{RS_0}+\frac{(\gamma+R)D_1}{RS_1}+1}
\end{align}

The population of the excited triplet state in the steady state is
\begin{align}
\begin{split}
N_e^{\text{st}} = N_3^{\text{st}} + N_4^{\text{st}} &= \left(\frac{D_0}{S_0}+\frac{D_1}{S_1}\right)N_5^{\text{st}}\\[7pt]
&=\displaystyle\frac{D_0S_1+D_1S_0}{S_0S_1+D_0S_1+D_1S_0}\;\frac{R}{\displaystyle\frac{S_0S_1(D_0+D_1)+\gamma(D_0S_1+D_1S_0)}{S_0S_1+D_0S_1+D_1S_0}+R}\\[7pt]
&=P_{e}\;\frac{R/R_{\text{sat}}}{1+R/R_{\text{sat}}}
\end{split}
\end{align}
where $P_e = (D_0S_1+D_1S_0)/(S_0S_1+D_0S_1+D_1S_0)$ is the maximum population of the excited triplet state in the limit of high optical pumping rate.
Because the optical pumping rate is proportional to the intensity we obtain the following relation
\begin{equation}
I/I_{\text{sat}} = R/R_{\text{sat}}
\end{equation}
and therefore, $R_{\text{sat}}$ can be understood as an optical pumping rate when the intensity is given as the saturation intensity $I_{\text{sat}}$ 
and it is given as
\begin{equation}
R_{\text{sat}}=\frac{S_0S_1(D_0+D_1)+\gamma(D_0S_1+D_1S_0)}{S_0S_1+D_0S_1+D_1S_0} = 27.38\;\text{MHz}.
\end{equation}

\section{Parameters Used in the Numerical Simulations}
\label{sec:numerical_parameters}

The parameters used in the Ramsey, spin-echo, and CW-ODMR simulations are summarized in Table~\ref{tab:simulation_parameters}. The analytical ensemble-sensitivity expression derived above employs the small-contrast approximation, \(C\ll1\), under which the photon-shot-noise is approximated
as \(\delta S_{\mathrm{ens}}\simeq\sqrt{Nn_{\mathrm{avg}}}\).
Because the numerical simulations use \(C=0.35\), as listed in Table~\ref{tab:simulation_parameters}, we assessed the quantitative effect of this approximation by recalculating the sensitivity gains using the
exact contrast-dependent photon-shot-noise expression.

For Ramsey magnetometry, the exact calculation yields a sensitivity gain of \(16.49~\mathrm{dB}\), compared with \(16.83~\mathrm{dB}\) in the small-contrast limit. For spin-echo magnetometry, the corresponding gains
are \(14.89~\mathrm{dB}\) and \(14.61~\mathrm{dB}\), respectively. Thus, using the exact photon-shot-noise expression changes the predicted gain by only \(0.34~\mathrm{dB}\) for Ramsey magnetometry and
\(0.28~\mathrm{dB}\) for spin-echo magnetometry. These results show that the small-contrast approximation has only a minor quantitative effect at \(C=0.35\) and does not affect the main conclusion regarding the
sensitivity enhancement achieved by selecting the optimal spin subset. All sensitivity-threshold and gain simulations presented in this work assume photon-shot-noise-limited detection \((\sigma_d=0)\). Detector noise is included only in the general sensitivity expression derived in Sec.~\ref{section:Detector Noise}.

\begin{table*}[ht]
\caption{\label{tab:simulation_parameters}
Parameters used in the Ramsey, spin-echo, and CW-ODMR simulations.}
\begin{ruledtabular}
\begin{tabular}{llcl}
Protocol & Parameter & Symbol & Value or implementation \\
\hline
Common
& Electron gyromagnetic  ratio
& $\gamma_e$
& $175.9\times10^{9}\ \mathrm{rad~s^{-1}~T^{-1}}$
\\[2pt]

Common & Inhomogeneous dephasing time
& $T_2^*$
& $0.45~\mu{\rm s}$~\cite{barry_sensitivity_2020, Schloss2018}

\\
 Ramsey / Spin-echo& Readout time& $t_{\text{readout}}$&0.242 $\mu {\rm s}$ (Sec.~\ref{section:five-level})\\
 Ramsey / Spin-echo& $m_s = 0$ spin state fluorescence& $a$&4.267 (Sec.~\ref{section:five-level})\\
 Ramsey / Spin-echo& $m_s=1$ spin state fluorescence& $b$&2.045 (Sec.~\ref{section:five-level})\\
 Ramsey / Spin-echo& contrast& $C$ & $(a-b)/(a+b)=0.35$ \\
 Ramsey / Spin-echo& mean number of detected photons & $n_{\text{avg}}$ & $(a+b)/2=3.16$\\

 Ramsey / Spin-echo
& Interrogation time
& $\tau_{\rm}$
& $T_2^*/2 = 0.225~\mu{\rm s}$

\\ Ramsey / Spin-echo
 & Antenna center Rabi frequency
 & $\Omega_0$
 & $57.22\times10^6 \: \mathrm{rad~s^{-1}}$
 
\\Spin-echo
& Echo coherence time
& $T_2$
& $7~\mu{\mathrm s}$~\cite{barry_sensitivity_2020, Schloss2018}

\\CW-ODMR
& Antenna center Rabi frequency
& $\Omega_{0}$
& $1.192 \times 10^6 \mathrm{rad~s^{-1}}$

\\CW-ODMR
& Antenna center optical saturation parameter
& $s_{0}$
& $0.03$

\\ CW-ODMR 
& Intrinsic population-relaxation rate 
& $\Gamma_1$
& $1\times10^3 \mathrm{s^{-1}}$~\cite{Dreau_avoiding}

\\ CW-ODMR 
& Inhomogeneous dephasing rate 
& $\Gamma_2^*$
& $1/T_2^* = 2.22\times10^6~\mathrm{s^{-1}}$

\\ CW-ODMR 
& Saturated optical-polarization rate 
& $\Gamma_p^\infty$
& $5\times10^6~\mathrm{s^{-1}}$~\cite{Dreau_avoiding}

\\ CW-ODMR 
& Saturated optical-cycle rate 
& $\Gamma_c^\infty$
& $8\times10^7~\mathrm{s^{-1}}$~\cite{Dreau_avoiding}

\\ CW-ODMR 
& $m_s=0$ spin state photon count rate 
& $r_a$
& $2.5\times10^5~\mathrm{counts\,s^{-1}}$ ~\cite{Dreau_avoiding}

\\ CW-ODMR 
& $m_s=1$ spin state photon count rate 
& $r_b$
& $0.6r_a=1.5\times10^5~\mathrm{counts\,s^{-1}}$
~\cite{Dreau_avoiding}

\end{tabular}
\end{ruledtabular}
\end{table*}

\section{HFSS for Rabi-frequency map}

We performed electromagnetic simulations using \textit{High Frequency Structure Simulator} (HFSS) to obtain the spatial distribution of the Rabi frequency generated by the circular loop microwave antenna. In our simulation, we assume that the quantization axes of the NV centers are aligned along the \([111]\) crystallographic direction, and the Rabi frequency is calculated accordingly. The diamond sample is modeled with dimensions \(5~\mathrm{mm} \times 5~\mathrm{mm} \times 0.5~\mathrm{mm}\), using the default \texttt{diamond} material in HFSS. The simulation resolution is set to \(500 \times 500\). The geometric parameters of the circular loop antenna used in the simulation are summarized in Table~\ref{tab:omega_dims}.

\begin{table}[ht]
\centering
\caption{Dimensions of the circular loop antenna and diamond used in the HFSS simulation.}
\label{tab:omega_dims}
\begin{tabular}{lc}
\hline\hline
Parameter & Value (mm) \\
\hline
Antenna radius
& $0.8$
\\

Trace width
& $0.127$
\\

Antenna thickness
& $0.018$
\\

Feed-line gap
& $0.2$
\\

Diamond lateral dimensions
& $5\times5$
\\

Diamond thickness
& $0.5$
\\
\hline\hline
\end{tabular}
\end{table}

\FloatBarrier
\section{Digital holography for structured illumination}

We use the \textit{Mixed-Region Amplitude Freedom} (MRAF) algorithm \cite{Pasienski_2008} to compute the hologram that efficiently generates structured illuminations. Similar to the Gerchberg-Saxton (GS) algorithm \cite{Kim2019}, the MRAF algorithm updates the hologram at the SLM plane based on the iterative two-dimensional Fourier and inverse Fourier transforms. The algorithm, unlike the GS algorithm, intentionally produces non-zero amplitude outside of the area of interest, and uses the redundancy to enhance the uniformity of the structured illuminations.

The initial $\phi_0 (x,y)$ of the MRAF algorithm is set to defocus the input Gaussian beam to make its beam diameter comparable to the target illumination pattern. After 300 iterations, the simulated non-uniformity saturates at 2.02 \%, and the resulting hologram is displayed on the SLM. In this calculation, the amplitude ratio of 55 \% is used for the redundancy. Because optical aberrations and residual SLM phase errors are not captured by the simulation, the realized illumination pattern deviates from the target. We further increase the uniformity of the structured illuminations using camera-based feedback \cite{Bruce2015}. After 19 iterations of the camera feedback loop, we achieved an intensity non-uniformity of 32.8 \% in the target region, with a corresponding power efficiency of 30.7 \%.

\bibliographystyle{apsrev4-2}
\bibliography{SuppReference}